\documentclass[sn-mathphys]{sn-jnl}

\usepackage{graphicx}%
\usepackage{multirow}%
\usepackage{amsmath,amssymb,amsfonts}%
\usepackage{amsthm}%
\usepackage{mathrsfs}%
\usepackage[title]{appendix}%
\usepackage{xcolor}%
\usepackage{textcomp}%
\usepackage{manyfoot}%
\usepackage{booktabs}%
\usepackage{algorithm}%
\usepackage{algorithmicx}%
\usepackage{algpseudocode}%
\usepackage{listings}%
\usepackage[utf8]{inputenc} 
\usepackage{hyperref}
\usepackage{breakurl}
\DeclareMathOperator{\diag}{diag}
\usepackage{amsmath}
\usepackage{amsthm}
\usepackage{amssymb}
\usepackage{dsfont}
\usepackage{graphicx}
\usepackage{cases}
\usepackage{balance}
\usepackage{here}

\DeclareMathOperator{\sgn}{sgn}
\DeclareMathOperator{\Sgn}{Sgn}
\usepackage{enumitem}
\usepackage{xcolor}
\definecolor{mygreen}{RGB}{0,120,90}

 	\definecolor{brightpink}{rgb}{1.0, 0.0, 0.5}
\usepackage{xcolor}       
\usepackage{todonotes}    
\newcommand{\R}{\mathbb{R}}
\definecolor{brightmaroon}{rgb}{0.76, 0.13, 0.28}
\usepackage[most]{tcolorbox} 
\usepackage{mathrsfs}
\usepackage{mdframed}
\newcommand{\ds}{\textbf{d}}
\theoremstyle{plain}
\newtheorem{theorem}{Theorem}
\newtheorem{maintheorem}{Main Theorem}

\newtheorem{lemma}{Lemma}
\newtheorem{remark}{Remark}
\newtheorem{definition}{Definition}

\newtheorem{assumption*}{Assumption}

\newtheorem{property}{Property}
\usepackage[english]{babel}

\DeclareMathOperator\sign{sgn}
\newcommand{\scal}[2]{\left\langle #1,\, #2\right\rangle  }
\newcommand{\norm}[1]{\left\Vert#1 \right\Vert}

\newenvironment{proofof}[1]{\par  \quad {\it Proof of #1:}}{\hfill $\blacksquare$\endtrivlist}

\theoremstyle{thmstyleone}%

\begin{document}

\title[On Robust Fixed-Time Stabilization of the Cauchy Problem in Hilbert Spaces]
{On Robust Fixed-Time Stabilization of the Cauchy Problem in Hilbert Spaces}


\author*[1]{\fnm{Moussa} \sur{Labbadi}}
\email{moussa.labbadi@enib.fr}

\author[2]{\fnm{Christophe} \sur{Roman}}

\author[3]{\fnm{Yacine} \sur{Chitour}}
\email{christophe.roman@lis-lab.fr,yacine.chitour@l2s.centralesupelec.fr}


\affil*[1]{\orgname{IRDL, UMR CNRS 6027, Bretagne INP, ENSTA, Institut
Polytechnique de Paris, Univ. Brest, Univ. Bretagne Sud},\orgaddress{\city{Brest}, \country{France}}} 

\affil[2]{
\orgname{Aix-Marseille Univ, CNRS, LIS}, 
\orgaddress{\city{Marseille}, \country{France}}}

\affil[3]{
\orgname{Laboratoire des Signaux et Syst\`emes (L2S), Universit\'e Paris-Saclay, CNRS, CentraleSup\'elec}
\orgaddress{\city{Gif-sur-Yvette}, \country{France}}}

\abstract{This paper presents finite-time and fixed-time stabilization results for inhomogeneous abstract evolution problems. We prove well-posedness for strong and weak solutions, and estimate upper bounds for settling times for both homogeneous and inhomogeneous systems. We generalize finite-dimensional results to infinite-dimensional systems and demonstrate partial state stabilization with actuation on a subset of the domain. The interest of these results are illustrated through an application of a heat equation with memory term.}

\keywords{Finite-/ Fixed-Time Stability, Nonlinear Feedback Control, Maximal Monotone Operators, Robust Stabilization.}



\maketitle

\section{Introduction}
For evolution problems and boundary control, finite-time convergence can be framed as follows: under which conditions does a solution of an evolution problem have compact time support (i.e., the solution becomes zero after a finite time). Such phenomena are often linked to discontinuous or non-differentiable dynamics, notably involving the ``$\sgn$'' operator. Several mathematical approaches come together in addressing this problem and we next list a few. The first one deals with maximal monotone operator theory \cite{Brezis1968,Brezis1973}, where it is shown that differential inclusions of the type
\(
\dot u + \partial \Phi(u)\ni~0
\)
may exhibit finite-time extinction due to strong dissipativity of subdifferentials, cf.  
\cite{diaz2005special} for instance. The second approach which concerns discontinuous dynamical systems theory (as presented in \cite{Filippov1964} and extended in \cite{Utkin1977}), underpins sliding-mode control for finite-time convergence and robustness. A third approach treats finite-time stability via Lyapunov function techniques especially when dynamics involves terms such as the set-valued map $\sgn(x)$ or $|x|^\alpha \sgn(x)$ with $\alpha\geq 0$, \cite{Roxin1966,BhatBernstein2000}. 

Finally a fourth and more modern approach considers homogeneous methods (with homogeneous dynamics) and it allows for the design of finite- or fixed-time stabilizers for evolution equations within Hilbert spaces, \cite{polyakov2018homogeneous,10742482,polyakov2021input}.

Homogeneity-based stabilization has been proposed as an effective control strategy to ensure robustness as well as finite- and fixed-time convergence, originating from the theory of homogeneous stabilization of nonlinear ordinary differential equations (ODEs), cf \cite{Polyakov2025} for historical remarks and complete definitions. In finite dimension, the core idea is to design a feedback law such that the resulting closed-loop system on $\mathbb{R}^n$ given by $\dot x=f(x)$ admits a homogeneity property defined as follows. Given a real $\mu$ and $r_1,\cdots,r_n$ positive real numbers, one says that the vector field \(f: \mathbb{R}^n \to \mathbb{R}^n \) (or the ODE $\dot x=f(x)$) is \( \ds \)-homogeneous of degree $\mu$ if $f(\ds(s)x) = e^{\mu s}\ds(s)f(x)$ for every $s \in \R$ and $x \in \R^n$, 
where \( \ds \) denotes the diagonal matrix given by $\diag(e^{sr_1},\cdots,e^{sr_n})$. The homogeneity degree $\mu$ of the closed-loop vector field $f$ plays a central role in characterizing possible convergence properties. To see that, consider a positive definite function $V:\R^n\to \R_+$ (tending to infinity as $x$ tends to infinity) which is \( \ds \)-homogeneous of degree $m>0$ (i.e., $V(\ds(s)x) = e^{m s}V(x)$ for every $s \in \R$ and $x \in \R^n$) so that 
for every trajectory $x(\cdot)$ of the closed-loop system, $\dot V\leq -cV^{1+\frac{\mu}m}(x(t))$, where $\dot V$ denotes the time derivative of $V$ along $x(\cdot)$ and $c>0$ is a constant. Then 
$\dot x=f(x)$ is globally asymptotically stable with respect to the origin where the rate of convergence of $t\mapsto V(x(t))$ 
depends on $\mu$: if $\mu>0$, $V(x(\cdot))$ tend to the origin at least as fast as $t^{-1/\mu}$, while for $\mu=0$ the convergence is at least exponential and if $\frac{\mu}m\in [-1,0)$ the convergence to the origin occurs in finite time. 

Homogeneity also ensures robustness to locally Lipschitz disturbances, as the scaling structure of the closed-loop vector field is preserved. Note also that these methods introduce non-Lipschitz right-hand sides for negative $\mu$ and even discontinuous ones for $\mu=-1$. Although \cite{10742482} motivates the approach using heat and wave equations, the analysis addresses linear PDEs with nonlinearities which are (essentially) homogeneous 
at the origin. In this reference, the 
well-posedness issue of the closed-loop equations for $\mu<0$ is not addressed, even for weak solutions. Moreover, the homogeneous Lyapunov function is defined implicitly, which leads to two fundamental limitations. First, the existence of strong solutions in the presence of matched perturbations is not rigorously established. Second, the sampled-time implementation of the implicit Lyapunov-function–based control guarantees only asymptotic stability, since the homogeneous norm is computed implicitly by the formula~(12) in \cite{10742482} (requiring the development of a numerical method for consistent discretization of the closed-loop system);
see also \cite{polyakov2015finite} and \cite{labbadi2025discretization} for further details). As a consequence, the resulting closed-loop properties do not strictly coincide with those obtained under an explicit Lyapunov function formulation, and the claimed finite- or fixed-time convergence may be lost in practical implementation \cite{labbadi2025discretization}.

Discontinuity and non-Lipschitz character at the origin for infinite-dimensional systems are  essential both for giving a rigorous meaning to solutions at all times and for achieving disturbance rejection. The reference \cite{brezis1974monotone} explicitly formulates this issue and provides sufficient conditions for inhomogeneous evolution problems to be stable at the origin: for instance it is asked that \( f(t) \in A0\ni\{0\} \), where $A0$ denotes the value at $0$ of a set-valued operator $A$
for the the differential inclusion
\(
\dot{u}(t) + A u(t) \ni f(t),
\) 
in order to have trajectories $u$ stable at the origin (adding of course stability assumption on $A$). If moreover, one wants that these trajectories reach the origin and possibly stay there, then this observation is (essentially) a necessary condition when considering general disturbances $f$. The above observation highlights a delicate 
issue: robust rejection of arbitrary bounded disturbances and finite-time convergence require discontinuous dynamics since the latter is usually modeled by nontrivial set-valued maps.

In control theory, the concepts of disturbance rejection, discontinuity and finite-time convergence are encompassed in sliding-mode control (SMC), as emphasized in  \cite{utkin2013sliding} for instance. For infinite-dimensional systems, SMC applies either through boundary control or distributed control. Boundary SMC restricts the sliding action to the spatial boundary, making the analysis technically demanding.

Early works established the theoretical foundations of SMC for infinite-dimensional systems \cite{orlov1987sliding,utkin1990control,levaggi2002infinite,orlov2002discontinuous,Orlov2025SlidingModeBoundaryControl}, focusing on discontinuous feedback laws and boundary control strategies \cite{orlov2002discontinuous,cheng2011sliding,colli2019sliding,levaggi2007regularization,levaggi2013existence,baji2007asymptotics}, later extended to linear and semilinear PDEs, robust output-feedback designs, and higher-order sliding algorithms such as super-twisting \cite{orlov2004robust,balogoun2025sliding,orlov2020nonsmooth,balogoun2022super}. Applications in chemical reactors, phase-field models, and tumor-growth dynamics further illustrate these approaches \cite{orlov2002discontinuous,colli2019sliding}, and recent developments have analyzed sliding motions under parameter variations and boundary uncertainties in parabolic and hyperbolic systems using rigorous Lyapunov and nonsmooth-analytic frameworks \cite{levaggi2002infinite,orlov2020nonsmooth,balogoun2025sliding}.

Most existing PDE-based SMC designs, particularly those inspired by \cite{levaggi2002infinite,orlov2020nonsmooth,balogoun2025sliding,orlov2004robust,balogoun2022super,levaggi2007regularization,levaggi2013existence,baji2007asymptotics}, rely on feedback laws that are either continuous but not $C^{1}$ at the origin, or fully discontinuous. This nonsmoothness raises delicate issues concerning the definition, existence, and well-posedness of solutions, especially in the presence of boundary disturbances. 
Sliding-mode motions inherently require discontinuity to ensure disturbance rejection, and nonsmooth analysis is challenging even in finite dimension, where several notions of solutions exist, including those of Carathéodory, Hermes, Krasovskii, and Filippov \cite{hajek1979discontinuous,orlov2020nonsmooth}. In finite-dimensional systems, the existence of solutions in sliding-mode systems typically relies on Filippov-type solutions, i.e., for an equation $\dot x=f(x)$,
\[
    \dot x \in \bigcap_{\varepsilon>0}\bigcap_{\small\ell(M)=0}\overline{\text{conv}} f((x+\varepsilon \mathcal B)\backslash M),
\]
where $M$ is the set of discontinuity points of $f$, $\ell$ is the Lebesgue measure and $\mathcal B$ denotes the unit ball of the state space. A primary challenge is the possible non-uniqueness of such solutions \cite{danca2010uniqueness,jeffrey2014dynamics}. Regularization techniques \cite{orlov2020nonsmooth} are often employed to handle discontinuities, but they may not yield Filippov-type solutions \cite{dieci2013regularizing}. Non-uniqueness can be incorporated into convergence analysis by considering all admissible solutions, although this requires non trivial computations typically using Lyapunov-based techniques.

For infinite-dimensional systems, the situation becomes even more subtle: regularization techniques usually only guarantee the existence of weak solutions, whereas Lyapunov-based analysis requires strong solutions, see for instance \cite{10742482,vanspranghe2021velocity,hayat2019quadratic}.
 Even the precise meaning of the ``sign'' relation must be specified: one may consider the real sign relation acting pointwise on scalars or a function sign function (see later in the notations section). We next consider sliding-mode differential inclusions of the type $\dot X(t) + A X(t) \ni f(t)$ where $A$ denotes a relation and $f$ can be either a feedback, a source term or a disturbance (see below for precise definitions). For these systems, three frameworks exist to handle discontinuous, sign-type nonlinearities. Regularization method by viability-theoretic methods \cite{levaggi2002sliding,levaggi2004high,aubin1982differential,vrabie1995compactness,carjua1997some,carjua2001viability} which yield mild solutions of the differenetial inclusion via tangency conditions to viability kernels. However, these do not ensure strong solutions nor invariance of $\mathcal{D}(A)$, as mild solutions may fail to satisfy a.e. $X(t)\in \mathcal{D}(A)$, and are quite technical in practice and restrictive. Typically, they demand compact semigroups, which exclude hyperbolic types of equations. Regularization approaches used in \cite{orlov2009discontinuous,orlov2020nonsmooth}, which approximate the sign discontinuity by a continuous function such as $\frac{u}{\|u\|_2+\epsilon}$ , or $\frac{u_t}{\|u_t\|_2+\epsilon}$ for $\epsilon > 0$ small enough.
Most of the results obtained following that technique (starting from the late nineties until \cite{10742482}) use a concept of ``generalized solution'' given in \cite[Definition~2.10]{orlov2009discontinuous} and refer to \cite[Theorem 2.4]{orlov2009discontinuous}, when the existence of such a solution is claimed. The main issue is that, despite its statement, the above theorem does not prove the existence of a generalized solution (which is actually a strong solution). A careful reading of what is presented as a proof of that theorem simply aims at showing that, if a generalized solution exists, then it is the limit of any approximating sequence made of regularized solutions.
One would rather try to prove that such approximating sequences made of regularized solutions in the sense of \cite[Definition~2.9]{orlov2009discontinuous} are Cauchy with respect to an appropriate topology, or at least to prove the existence of a converging subsequence. To the best of the authors’ knowledge, there is no proof of these facts at the present time. Moreover, even with the existence of a limit for the regularizing scheme (up to a subsequence), it must still be established that this limit is indeed a solution (in some sense) to the abstract evolution problem. 
As a consequence, it is necessary to bring clarification to the several interesting designs proposed in the literature which rely on \cite{orlov2009discontinuous,orlov2020nonsmooth} for existence of solutions. For instance,  the feedbacks proposed in \cite{pisano2011tracking} for distributed sliding-mode control for heat PDEs
\begin{align}\label{eq:heat}
    u_t = u_{xx} + Q(t,u) + d(t),
\end{align}
lack a proper treatment of well-posedness, even in the disturbance-free case ($d(t)\equiv 0$) with power-type SMC
\[
Q(t,u) = -k_1 \frac{u|u|^\alpha}{\|u\|_2}, \quad \alpha\in(0,1),\ k_1>0.
\]
Note that global finite time stability is achieved under the condition that strong solutions exist. As shown in the present paper, the classical theory of maximal monotone operators in Hilbert space allows one to establish the existence of strong solutions, and with some adjustments, can achieve fixed-time convergence. Note, nevertheless, that for super-twisting-like controllers, 
\[
Q(t,u) = -k_1 |u|^{1/2}\sgn(u) - k_2 u + \eta, \quad \eta_t = -k_3 \frac{u}{\|u\|_2} - k_4 u,
\]
(where the constant $k_i$'s have appropriate sign) we are not able to provide proper proofs of existence of weak or strong solutions. The same fact holds regarding the differential inclusion
\begin{align}\label{SM1partial}
    u_{tt} = u_{xx} + Q(t,u) + d(t),
\end{align}
with second-order sliding laws
\[
Q(t,u) = -k_1 \frac{u}{\|u\|_2} - k_2 \frac{u_t}{\|u_t\|_2}, \quad k_2 > \bar d, \quad k_1>k_2+\bar d.
\]
Note that, also in the above case, Lyapunov-type stability analysis crucially relies on the existence of  strong solutions. Moreover, in the case where $Q(t,u) = -k_1 \operatorname{sign}(u_t)$ and $d(t)\equiv 0$ in \eqref{SM1partial}, solutions may fail to converge to zero, cf. \cite{baji2007asymptotics}. Finally, it has to be reported that some papers simply assume the existence of (weak or strong) solutions and proceed with the design of controllers, cf. \cite{orlov2011boundary,pisano2012boundary}, while other papers provide ad hoc arguments for the existence of solutions especially in the specific case of boundary sliding mode \cite{liard2022boundary,balogoun2022slidingmodecontrolclass,chitour2021one,guzman2025rapid}.

In the present paper we give sufficient conditions for the existence of strong solution and fixed time convergence, despite bounded disturbances associated with abstract formulation $\frac{d}{dt}X(t)+AX(t)\ni f(t)$. Using maximal monotone operator theory and contraction semigroups in Hilbert spaces \cite{brezis1973ope,barbu1993analysis}, we obtain Carathéodory-type solutions with right-hand derivatives everywhere. The paper further addresses an intermediate regime between first-order and second-order sliding modes. Our main contributions are summarized as follows: we first
establish well-posedness for an evolution problem governed by a maximal monotone operator, guaranteeing the existence of strong and weak solutions and a finite extinction time. We then extend that result to a 
 a subspace defined as the null space of a linear relation, under specific assumptions, again ensuring well-posedness and finite extinction time. We finally provide a variation of Main Theorem 2 under alternative assumptions. These findings 
are illustrated with examples based on the heat equation with and without memory terms. 

The remainder of this paper is organized as follows. Section 2 introduces the notations and preliminary results used throughout the paper. Section~3 presents the main theoretical results, including both the full state and partial state fixed-time convergence results under uniform bounds. Section 4 provides the proof of the main results. Section 5 is devoted to an application to the heat equation with modal projection, including the problem formulation, verification of assumptions, and the associated control law and convergence analysis. Section 6 discusses additional extensions and remarks. Section 7 presents a counterexample showing that boundedness alone is insufficient for the existence of strong solutions. Finally, Section 8 concludes the paper and outlines possible directions for future work, including extensions to systems with memory effects and nonlinear coupling terms.

\section{Notations and Preliminaries}
\label{sec:problem}
Throughout this paper, $\mathtt{H}$ denotes a Hilbert space endowed with an inner product $\scal{\cdot}{\cdot}$
and the associated norm $\norm{\cdot}$.
In what follows we will consider relations on $\mathtt{H}$, i.e.,  set-valued maps on $\mathtt{H}$, not necessarily linear, cf. the notion of linear relations in \cite{cross1998multivalued}. In case the relation is single-valued we will refer to it as operator. \footnote{This terminology is not universal, since many authors use the term operator also for set-valued maps (or only for linear maps).} 


Otherwise mentioned, definitions, notations and properties on relations given below are taken from \cite{brezis1973ope}. A relation $G: \mathtt{H} \to 2^\mathtt{H}$ is identified with its graph $\mathcal{G}(G)$ in $\mathtt{H}\times \mathtt{H}$ and defined as $Gx=\{y:(x,y)\in \mathcal{G}(G)\} $. The domain and the range of a relation are equal to $\mathcal{D}(G)=\{ x: Gx\neq \varnothing \}$ and $\mathcal{R}(G)=\bigcup_{x\in\mathcal{D}(G)} Gx$ respectively. The inverse of a relation is defined as the graph $(y,x)\in \mathcal{G}(G^{-1}) \Leftrightarrow (x,y)\in \mathcal{G}(A)$. An extension of a relation $G$ is a relation containing $G$ with respect to set inclusion and a maximal extension of $G$ is an extension of $G$ not strictly contained in any extension of $G$. 
The set-valued map $G^\mathrm{o}$ denotes the minimal section of $G$, i.e.,  the map defined on $\mathcal{D}(G)$ which assigns to $x\in\mathcal{D}(G)$ the elements of least norm in $Gx$.

The pointwise set-valued sign map is defined by 
\[
\sgn : \mathbb{R} \to 2^\mathbb{R}, \quad 
\sgn(x) = 
\begin{cases}
\frac{x}{|x|} & \text{if } x \neq 0, \\ 
[-1,1] & \text{if } x = 0.
\end{cases}
\]
while the functional sign relation is given by 
\[
\Sgn : \mathbb{\mathtt{H}} \to 2^\mathtt{H}, \quad 
\Sgn(x) = 
\begin{cases}
\frac{x}{\norm{x}} & \text{if } x\neq 0,\\
\{z;\Vert z\Vert \leq 1\} & \text{otherwise}.
\end{cases}
\]
For $\alpha\geq 0$, we use $\left\lceil x \right\rfloor^\alpha$ to denote $\norm{x}^\alpha\Sgn(x)$ for $x\in \mathtt{H}$. Note that $(1+\alpha)\left\lceil x \right\rfloor^\alpha$ is equal to the 
gradient of $\norm{x}^{1+\alpha}$, 
understood as a subdifferential for $\alpha=0$. 

\begin{definition}
A relation $G$ on $\mathtt{H}$ is said to be monotone if for all  $(x,y)\in \mathcal{G}(G),  (x',y')\in \mathcal{G}(G)$ it holds
\begin{equation}
 \scal{y-y'}{x-x'}\geqslant 0,
\end{equation}
and is said to be maximal monotone if moreover $G$ admits no proper monotone extension. Additionally we denote $G\in \mathfrak{A}(\omega)$ with $\omega\in\mathbb{R}$ if $G+\omega I$ is maximal monotone. A relation $G$ is said to be dissipative if $-G$ is monotone and m-dissipative if it admits no proper dissipative extension.
\end{definition}

The main example of maximal monotone relation on $\mathtt{H}$ used in the paper is $x\mapsto \left\lceil x \right\rfloor^\alpha$ defined for any $\alpha\geq 0$, cf. \cite[Chapter 2]{brezis1973ope,baji2007asymptotics}.
$M$-dissipative relations are particularly important in the study of well-posedness, as they serve as the infinitesimal generators of strongly continuous contraction semigroups (cf. \cite{komura1967nonlinear} also for a definition).

\begin{property}
Consider $G$ a relation in $\mathtt{H}$. The following items are equivalent: $(a)$
$G$ is maximal monotone; $(b)$ $G$ is monotone and $\mathcal{R}(I+G)=\mathtt{H}$; $(c)$
$\forall \lambda >0$, $(I+\lambda G)^{-1}$ is a contraction defined on $\mathtt{H}$, i.e., $\forall x,y\in \mathtt{H}$ it holds
\begin{align}
    \norm{(I+\lambda G)^{-1}x-(I+\lambda G)^{-1}y}\leq \norm{x-y}.
\end{align}
\end{property}
\begin{definition}
    A relation $G$ is accretive if it holds $\forall \lambda>0, \forall (x,y),(x',y')\in\mathcal{G}(G)$
    \begin{equation}
 \norm{x-x'}\leqslant \norm{x-x'+\lambda(y-y')}.
\end{equation}
\end{definition}
\begin{property}
Consider $G$ a operator on $\mathtt{H}$. The following items are equivalent: 
$(i)$ $G$ is monotone; $(ii)$ $G$ is accretive; $(iii)$ $-G$ is dissipative.
\end{property}

Consider the following abstract problem
\begin{equation}\label{sys_abs}
    \frac{d X(t)}{dt}+AX(t)\ni f(t), 
\end{equation}
where $A$ is a  relation. We next provide the definitions of strong and weak solutions of \eqref{sys_abs} cf. \cite{brezis1973ope} or \cite{benilan1972solutions}.
\begin{definition}\label{def:sws}
Let $A$ a relation on $\mathtt{H}$ and $f\in L^1([0,T];\mathtt{H})$. 
\begin{itemize}
\item A strong solution of \eqref{sys_abs} is a function $u\in C^0([0,T],\mathtt{H})$ 
if for all $t\in[0,T]$ $u(t)\in\mathcal{D}(A)$ and for almost every $t\in [0,T]$ $\frac{d}{dt}u(t)+Au(t)\ni f(t)$.
\item A weak solution of \eqref{sys_abs} is a function $u\in C^0([0,T],\mathtt{H})$ if there exist a sequence $(f_n)$ in $L^1([0,T],\mathtt{H})$
and a sequence $(u_n)$ in $C^0([0,T],\mathtt{H})$ such that $u_n$ is a strong solution of $\frac{d}{dt}u+Au\ni f_n$ where $f_n\to f$ in $L^1([0,T],\mathtt{H})$ and $u_n\to u$ uniformly in $[0,T]$.
\end{itemize}
\end{definition}
The approach adopted here to define weak solutions of \eqref{sys_abs} relies on the definition of strong solutions. Other definitions rely instead on integral formulation (such as the Duhamel or Volterra integral in the linear case or Bénilan integral solution in the nonlinear case) or a variational expression (for the latter, in the linear setting, cf. \cite{ball1977strongly}).

For the existence of such solutions, we refer to  \cite{brezis1971monotonicity} or \cite{benilan1972solutions}, which also provide a concept of uniqueness by selecting among possible solutions ones with additional properties. 
\begin{theorem}[Theorem~21 and subsequent remark in \cite{brezis1971monotonicity}]\label{theo_maximal}
Let $A$ be a maximal monotone relation on a Hilbert space
$\mathtt{H}$ and $f \in W^{1,1}(0,T;\mathtt{H})$. Then for all $X_0 \in \mathcal{D}(A)$, there exists a unique $X(t): \mathbb{R}^+ \to \mathtt{H}$ such that 
\begin{itemize}
    \item $X(0)=X_0$, $\forall t\geq 0$, $X(t)\in \mathcal{D}(A)$,
    \item for a.e. $t\in[0,T)$, $\frac{d}{dt}X(t)+AX(t)\ni f(t)$, 
    \item $t\mapsto X(t)$ is differentiable from the right on $[0,T)$ and $\frac{d^+}{dt}X(t)+{(AX(t)-f(t))}^{\mathrm{o}}=0$,
    \item $|\frac{d^+}{dt}X(t)|\leq |\frac{d^+}{dt}X(0)|+\int_0^t|\frac{d f}{ds}(s)|ds$
\end{itemize} 

In addition, given $f,\hat f$ and $u_0,\hat u_0$, and the corresponding solution $u,\hat u$ it holds for $t\in [0,T]$, $  |u(t)-\hat u(t)| \leq |u_0-\hat u_0| + \int_0^t|f(s)-\hat f(s)|ds$. The mapping $(X_0,f)\mapsto X$ can be extended by continuity from $\overline{\mathcal{D}(A)}\times L^1(0,T;\mathtt{H})$ into $C(0,T;\overline{\mathcal{D}(A)})$.
\end{theorem}

\begin{remark}\label{rem:WS-solutions}

As an immediate consequence of the theorem, the maximal monotone relation $A$ can be associated with a strongly continuous semi-group of contractions $S(t)$ on $\overline{\mathcal{D}(A)}\subset \mathtt{H}$ which can be defined by the Hille-Yosida exponential formula
\begin{align}
    S(t)x=\lim_{n\to\infty} (I+\frac{t}{n}A)^{-n}x,
\end{align}
cf \cite{crandall1969semi}, \cite{brezis1971monotonicity} and \cite{komura1967nonlinear} \cite{crandall1971generation} for more details.
Note also that weak solutions in the above theorem have an equivalent characterization with respect to Definition~\ref{def:sws}: they can be defined by the Duhamel formula in the linear case, where, for $X_0\in\overline{\mathcal{D}(A)}$, and $T>0$, 
it holds for $t\in [0,T)$ and $X$ an integral solution,
\begin{align}
    X(t)=S(t)X_0+\int_0^tS(t-s)f(s)ds\in\overline{\mathcal{D}(A)}.
\end{align}
or by Bénilian integral formula in the nonlinear case \cite{takahashi1976convergence}, for every $s,t\in [0,T]$ such that $s\leq t$ and for every $(x,y)\in \mathcal{G}(A)$,
\begin{align}
    \norm{X(t)-x}^2-\norm{X(s)-x}^2\leq 2\int_s^t \scal{y}{X(\tau)-x}d\tau.
\end{align}
\end{remark}

\begin{remark}
    Even if $A$ is a set-valued map, the associated semigroups are single valued in the case where $-A$ is m-dissipative, cf. \cite{komura1967nonlinear}. 
\end{remark}

We next provide sufficient conditions used later in order to check  maximal monotonicity. We say that $F:D\to \mathcal{X}$ is hemicontinous if $\forall (x,y)\in D^2$ and $\forall z\in \mathcal{X}$ then it holds $\scal{F(x+t y)}{z}\to \scal{Fx}{z}$ as $t\to 0^+$, cf. Definition 11.2 in \cite{deimling2013nonlinear}. Then the following holds. 
\begin{theorem}[Theorem 1.3 \cite{barbu1993analysis}]\label{theo_bardu}
Let $\mathcal{X}$ be a reflexive Banach space and let $B:\mathcal{X}\to \mathcal{X}^*$ be a monotone hemicontinous operator. Then $B$ is maximal monotone in $\mathcal{X}\times \mathcal{X}^*$.
\end{theorem}
\begin{theorem}[Theorem~1.5 in \cite{barbu1993analysis} ]\label{theo_rockafellar}
   Let $\mathcal{X}$ be a reflexive Banach space  and let $A$ and $B$ be maximal monotone relations of $\mathcal{X}\times \mathcal{X}^*$. If 
   $(\text{int}\, \mathcal{D}(A))\cap \mathcal{D}(B)\neq \varnothing$,  then $A+B$ is maximal monotone.
\end{theorem}

We then recall some classic result from linear operator, cf. \cite{cross1998multivalued}.
\begin{definition}
    A linear operator $B$ from $X\to X$, (or a single-valued map) is said to be: non negative if $\forall x \in X$,  $\scal{x}{Bx}\geq 0$;
   self-adjoint if $\forall x,y\in X$, $\scal{x}{By}=\scal{Bx}{y}$; 
 projective if B$^2=B$. We use $N(B)$ to denote its kernel. 
 \end{definition}
 The following lemma is immediate to derive. 
 \begin{lemma}\label{lemma}
Consider that $B$ is a closed linear operator, which is also positive and self-adjoint.
    The bilinear product
    \begin{align}
        \scal{y}{x}_{\mathtt{H}/N(B)}=\scal{y}{Bx},
    \end{align}
    is a scalar product of the quotient Hilbert space ${\mathtt{H}/N(B)}$. The associated semi-norm in $\mathtt{H}$
    \begin{align}
        \norm{x}^2_{\mathtt{H}/N(B)}= \scal{x}{x}_{\mathtt{H}/N(B)},
    \end{align}
    is a norm of ${\mathtt{H}/N(B)}$.
\end{lemma}


  \begin{lemma}[\label{lemma1} Inspired from the results in finite dimension \cite{fxt:Polyakov:2012}]
Consider an Hilbert space $\mathtt{H}$, a positive definite functional $V:\mathtt{H}\to \mathbb{R}_+$ admitting a right Dini's derivative, and $x:\mathbb{R}_+\to\mathtt{H}$ admitting also a right Dini's derivative such that there exist $\beta_1,\beta_2>0$ and $c_2>1, c_1 \in (0,1)$ for which $\frac{d^+}{dt}V(x(t))\leq -\beta_1 V(x(t))^{c_1}-\beta_2V(x(t))^{c_2}$ 
for $t\geq 0$. Then $V(x(t))=0$ for $t\geq T_0(x(0)):= \frac{1}{\beta_1(1-c_1)}+\frac{1}{\beta_2(c_2-1)}$.
\end{lemma}
\section{Main results} 
\label{sec:controller}
In this section, we present the main results of the paper. We establish the necessary conditions for the existence and uniqueness of both strong and weak solutions to a given abstract evolution problem in a Hilbert space, ensuring that the length of the support has a uniform upper bound independent of the initial condition. Furthermore, we provide the necessary conditions for extending this result to partial state stabilization rather than the full state. The last section presents the application of the final result to a specific partial differential equation.

Consider the following  abstract control problem
 for $t\in [0,T]$
\begin{subnumcases}{\label{sys_control}}
\frac{d}{dt} X(t)+AX(t)+ UX(t)\ni d(t)   \\
X(0) = X_0,   
\end{subnumcases}
where both $A$ and $U$ are relations and $ d\in L^1(0,T;\mathtt{H})$.
Our objective is to design a control law \( U: \mathtt{H} \to 2^{\mathtt{H}} \) such that the solution \( X \) is well-posed and has compact support.

\subsection{Full state fixed time
}
Let us consider the following inhomogenous problem
\begin{subnumcases}{\label{sys_ab_inhomo_fixed}}
    \frac{d}{dt} X(t)+AX(t)+ k\lceil X(t)\rfloor^0 +k_2 \lceil X(t)\rfloor^{\alpha}\ni d(t),   \\
    X(0) = X_0.   
    \end{subnumcases}

\begin{maintheorem}\label{theo_inhomo_fixed}
    If $k>0$, $k_2>0$ and $\alpha>1$, the abstract evolution problem \eqref{sys_ab_inhomo_fixed} is well-posed in the case where $A:\mathtt{H} \to 2^\mathtt{H}$ is maximal monotone on the Hilbert $\mathtt{H}$, for all $T>0$,
    \begin{itemize}
        \item if $d\in W^{1,1}(0,T;\mathtt{H})$ and the existence and uniqueness of strong solution holds, i.e., $\forall X_0\in\mathcal{D}(A)$ ,
        \begin{align}
          X\in C^0([0,T);\mathcal{D}(A)) ,\  \frac{d}{dt} X\in L^\infty([0,T);\mathtt{H}).\label{sol_strong_2}
        \end{align}
        \item if $d\in L^{1}(0,T;\mathtt{H})$ and $\forall X_0\in \overline{\mathcal{D}(A)}$, then the existence and uniqueness of weak solution holds on $\overline{\mathcal{D}(A)}$, i.e.,
        \begin{align}
            X\in C^0([0,T); \overline{\mathcal{D}(A)}).\label{sol_weak_2}
        \end{align}
    \end{itemize}
    If in addition $d\in L^\infty$, with $\|d(t)\|_{L_\infty}<\bar{d}$, $0\in A0$ and $k>\bar{d}$, then there exists $T_0(X_0)> 0$ such that, for solutions defined in   \eqref{sol_strong_2} and  \eqref{sol_weak_2}, one has
    \begin{align}
        X(t)=0,\quad \forall t\geq T_0(X_0).
    \end{align}
    where 
    \begin{align}\label{settlingtime_FxT}  
        T(X_0)\leq   T_{\max} = \frac{1}{k - \bar d}+\frac{1}{k_2 (\alpha - 1)}.
        \end{align}
\end{maintheorem} 

\begin{remark}\label{rem:mainth1}
In the special case $k_2 = 0$ and $d = 0$, this result has been established for weak solutions in a Banach space $\mathtt{B}$ whose dual $\mathtt{B}^*$ is convex, a setting that includes Hilbert spaces in \cite{barbu1993analysis} Proposition~2.3. Furthermore, the proof of finite-time convergence does not rely on the Lyapunov method.
\end{remark}

\subsection{Partial state fixed time
}
In this section, we provide several conditions for achieving partial finite-time stability. First, we analyze the system under feed-forward compensation to mitigate disturbances. Next, we address the scenario with a bounded uncontrolled part, offering solutions to ensure stability within a finite time. We consider two cases. In the first one, we directly compensate the influence of the uncontrolled part on the controlled part using a feedforward term. In the second case, the result relies on the fact that the influence of the uncontrolled part can be bounded by the controlled part, making direct compensation avoidable under the bounded hypothesis.

Let us consider the following abstract problem
\begin{subnumcases}{\label{sys_partial}}
\frac{d}{dt} X(t)+A X(t)+k_2 \lceil BX(t)\rfloor^{\alpha} +k\lceil BX(t)\rfloor^0-BAX(t)\ni Bd(t),\\
X(0) = X_0,    
\end{subnumcases}
As we examine the conditions for compact support while assuming $ d(t) = 0 $ for all $ t \geq T $, we require that $ d(t) \in A0 \cup B k\sign(0) $. Since our goal is to impose assumptions only on $ k $ without overly constraining $ A $, we choose to restrict the inhomogeneous term through the linear operator $ B $.

\begin{maintheorem}\label{theo_inhomo_fixed_partial_feed}
    If $k>0$, $k_2>0$ and $\alpha>1$, the abstract evolution problem \eqref{sys_partial} is well-posed in the case where $(I-B)A:\mathtt{H} \to 2^\mathtt{H}$ is maximal monotone on the Hilbert $\mathtt{H}$ and the linear closed operator $B$ is self-adjoint and projective. Given  $T>0$, 
    \begin{itemize}
        \item if $Bd\in W^{1,1}(0,T;\mathtt{H})$ it holds the existence and uniqueness of strong solution, i.e., $\forall X_0\in\mathcal{D}(A)$ ,
        \begin{align}
          X\in C^0([0,T);\mathcal{D}(A)) ,\  \dot X\in L^\infty([0,T);\mathtt{H}).\label{sol_strong_3}
        \end{align}
        \item if $Bd\in L^{1}(0,T;\mathtt{H})$ and $\forall X_0\in \overline{\mathcal{D}(A)}$, then the existence and uniqueness of weak solution holds on $\overline{\mathcal{D}(A)}$, i.e., 
        \begin{align}
            X\in C^0([0,T); \overline{\mathcal{D}(A)}).\label{sol_weak_3}
        \end{align}
    \end{itemize}
    If in addition $Bd\in L^\infty$, with $\forall t$, $B|d(t)|\leq \bar{d}$, $0\in A0$ and $k>\bar{d}$, then it exists $T_0(X_0)> 0$ such that, for solutions defined in \eqref{sol_strong_3} and \eqref{sol_weak_3}, one has 
    \begin{align}
        \norm{X(t)}_{X/N(B)}=0,\quad \forall t\geq T_0(X_0).\label{eq_fixed}
    \end{align}
    where 
    \begin{align}\label{settlingtime_FxT_2}  
        T_0(X_0)\leq   T_{\max} = \frac{1}{k - \bar d}+\frac{1}{k_2 (\alpha - 1)} .
        \end{align}
\end{maintheorem} 

We can consider the system \eqref{sys_partial} as a dynamical plant model by $A$ with the following feedback law
\begin{align}\label{feedback_forward_law}
  UX(t)=& -k_2 \lceil BX(t)\rfloor^\alpha-k\lceil BX(t)\rfloor^0-BAX(t).
\end{align}
Here, the use of $ B $ highlights that the control acts only on a part of the state. The computations performed above remain applicable, ensuring fixed-time stabilization of the null space of $B$ ($N(B)=B^{-1}(0)$). This extension is of practical interest. In other words, we look at the equation on the Hilbert space defined as the quotient space $\mathtt{H}/N(B)$. The feedforward term is given by $BAX(t)$ but we can also take
\begin{align}
 BX(t)\frac{\scal{AX(t)}{BX(t)}}{\norm{X}_{\mathtt{H}/N(B)}^2} ,  \label{amakali}
\end{align}
 as feedforward term as in \cite{amaliki_2024}.

\begin{remark}\label{rem:mainth2}
The fixed-time convergence result \eqref{eq_fixed} implies convergence on the quotient space $X / N(B)$. In particular, the smaller the kernel of $B$, the stronger the effect of the result in guiding the trajectory. The limiting cases are $B = 0$, where $N(B) = X$ and no convergence is observed, and $B = I$, where convergence over the entire space is achieved. In the latter case, note that $BAX = AX$, so the feedforward term exactly compensates the operator $A$, rendering the result trivial. For this reason, we investigate solutions without the $-BAX$ term.
\end{remark}

In the sequel, we no longer require the boundedness of the uncontrolled part in order to establish finite-time convergence and perturbation rejection. In the context of feedback stabilization, we consider the feedback law
\begin{align}\label{feedback_law}
    UX(t)=& -k_2 \lceil BX(t)\rfloor^\alpha-k\lceil BX(t)\rfloor^0,
 \end{align}
so the closed loop system is given by
\begin{subnumcases}{\label{sys_partial_2}}
    \frac{d}{dt} X(t)+A X(t)+k_2 \lceil BX(t)\rfloor^\alpha+k\lceil BX(t)\rfloor^0 \ni Bd(t),\\
    X(0) = X_0.    
\end{subnumcases}
We show the following result.
\begin{maintheorem}\label{theo_partial_2}
  If $k>0$, $k_2>0$ and $\alpha>1$, the abstract evolution problem \eqref{sys_partial_2} is well-posed in the case where $A:\mathtt{H} \to 2^\mathtt{H}$ is maximal monotone on the Hilbert $\mathtt{H}$, the linear closed operator $B$ is positive, self-adjoint and projective. Given $T>0$,
    \begin{itemize}
        \item if $Bd\in W^{1,1}(0,T;\mathtt{H})$ then the existence and uniqueness of strong solution holds, i.e., $\forall X_0\in\mathcal{D}(A)$ ,
        \begin{align}
          X\in C^0([0,T);\mathcal{D}(A)) ,\  \dot X\in L^\infty([0,T);\mathtt{H}).\label{sol_strong_4}
        \end{align}
        \item if $Bd\in L^{1}(0,T;\mathtt{H})$ and $\forall X_0\in \overline{\mathcal{D}(A)}$, then the existence and uniqueness of weak solution holds on $\overline{\mathcal{D}(A)}$, i.e., 
        \begin{align}
            X\in C^0([0,T); \overline{\mathcal{D}(A)}).\label{sol_weak_4}
        \end{align}
    \end{itemize}
    If in addition $Bd\in L^\infty$, with $\forall t, \ B \|d(t)\|<\bar{d}$, $0\in A0$ and $k-\omega_1>\bar{d}$, and with $\forall x\in\mathcal{D}(A),\  y\in Ax$ 
    \begin{align}
  \scal{y}{Bx}\geq -\omega_1\norm{Bx}-\omega_2\norm{Bx}^{\alpha+1}  \label{ABx}.
    \end{align} then there exist $T_0(X_0)> 0$ such that, for solutions defined in \eqref{sol_strong_4} and \eqref{sol_weak_4}, one has
    \begin{align}
        \norm{X(t)}_{X/N(B)}=0,\quad \forall t\geq T_0(X_0).
    \end{align}
    where 
    \begin{align}\label{settlingtime_FxT_3}  
        T_0(X_0)\leq  T_{\max}=\frac{1}{k-\omega_1 - \bar {d}}+\frac{1}{(k_2-\omega_2) (\alpha - 1)}.
        \end{align}
\end{maintheorem}

    The requirement in Main Theorem~\ref{theo_inhomo_fixed_partial_feed} for $ (I-B)A$ to be maximal monotone imposes a significant constraint on the feedforward control law. In \cite{amaliki_2024}, with the use of the feedforward term \eqref{amakali}, the well-posedness was established by leveraging the fact that this term is Lipschitz and associated with $ BA - cI $ being monotone (for some real $c$). In Theorem~\ref{theo_partial_2}, the condition shifts to requiring that the uncontrolled part can be bounded by the controlled part, which, while also a stringent assumption, represents a different type of constraint. Indeed using $A$ monotone one gets, $\forall x\in \mathcal{D}(A)$, $y\in Ax$
    \begin{align}
        \scal{y}{Bx}\geq -\scal{y}{x-Bx}.
    \end{align}
    Therefore  
    \begin{align}
        \scal{y}{x-Bx}\leq \omega_1\norm{Bx}+\omega_2\norm{Bx}^{\alpha+1},
    \end{align}
    is a sufficient condition for \eqref{ABx} to holds.

\section{Proof of the mains results}

We need the following technical result. 

\begin{lemma}\label{lem_par}
    Consider $B$ is a linear closed operator, with is self-adjoint and projective. Then the nonlinear relation $\mathtt{H}\ni x\to \lceil Bx \rfloor^\alpha$, $\alpha\geq 0$, is maximal monotone in $\mathtt{H}$.
  \end{lemma}

  \begin{proofof}{Lemma~\ref{lem_par}}
Recall first that for  $x\in\mathtt{H}$, one gets with $z\in \lceil x \rfloor^\alpha$, and $q\in \lceil x \rfloor^0$
  \begin{align}
      \scal{x}{z}=\scal{x}{q}\norm{x}^\alpha=\norm{x}^{\alpha+1}.
  \end{align}
For every $ x_1,x_2\in\mathtt{H}$,  let $y_1\in \lceil x_1 \rfloor^\alpha$, $y_2\in \lceil x_2 \rfloor^\alpha$, $z_1\in \lceil x_1\rfloor^0$, and $z_2\in \lceil x_2 \rfloor^0$, one has
  \begin{align}
  &\scal{x_1-x_2}{y_1-y_2}=\scal{x_1-x_2}{\Vert x_1\Vert z_1-\Vert x_2\Vert z_2}=\notag\\&\norm{x_1}^{\alpha+1}+\norm{x_2}^{\alpha+1}-\scal{x_1}{z_2}\norm{x_2}^{\alpha}-\scal{x_2}{z_1}\norm{x_1}^{\alpha}\geq\notag\\&
  \norm{x_1}^{\alpha+1}+\norm{x_2}^{\alpha+1}-\norm{x_1}\norm{x_2}^{\alpha}-\norm{x_2}\norm{x_1}^{\alpha}=\notag\\&
  (\norm{x_1}^\alpha-\norm{x_2}^\alpha)(\norm{x_1}-\norm{x_2})\geq 0,
  \end{align}
  which implies that the relation $x\to \lceil x \rfloor^\alpha, \alpha\geq 0$ is monotone. This is also the case for the relation $x\to \lceil Bx \rfloor^\alpha, \alpha\geq 0$, since $B$ is projective and self-adjoint, and  with $y_1\in \lceil x_1 \rfloor^\alpha$, and $y_2\in \lceil x_2 \rfloor^\alpha$, one gets
  \begin{align}
    \scal{x_1-x_2}{y_1-y_2}=\scal{Bx-By}{y_1-y_2}\geq 0.
    \end{align}
For establishing that $\mathtt{H}\ni x\to \lceil B x \rfloor^\alpha$ is maximal in the set of monotone operators we must check the range condition, i.e., $\forall  f\in\mathtt{H},\ \exists y\in \mathcal{D}(\lceil B \cdot \rfloor^\alpha)=\mathtt{H}$ such that
      \begin{equation}\label{eq:max-mon}
          y+\lceil B y \rfloor^\alpha\ni f.
      \end{equation}
For $\alpha=0$ and $f\in B^{-1}(0)$, take $y=f$ which is solution of \eqref{eq:max-mon} since $Bf=0$. Otherwise consider $y=f-\lceil B f \rfloor^\alpha$, which implies that 
\[
f=y+\lceil B f \rfloor^\alpha.
\]
Notice first that $By=(1-\frac1{\|Bf\|^{\alpha-1}})Bf$ since $B$ is projective. This implies that 
$\lceil B f \rfloor^\alpha\subset \lceil B y \rfloor^\alpha$ (with equality if $\alpha>0$), yielding that $y$ is solution of \eqref{eq:max-mon}.

  \end{proofof}

  Using the previous lemma we can established Theorem~\ref{theo_inhomo_fixed_partial_feed}.
  \begin{proofof}{Main Theorem~\ref{theo_inhomo_fixed_partial_feed}}
    Let $(I-B)A:\mathtt{H}\to 2^\mathtt{H}$ be a maximal monotone relation. Using Theorem~1.5 in \cite{barbu1993analysis} and Lemma~\ref{lem_par}, we establish that the relation $x\to Ax-BAx + k\lceil B x \rfloor^0 + k_2\lceil B x \rfloor^\alpha$ is also maximal monotone. Finally, Theorem~21  in \cite{brezis1971monotonicity} gives us the existence and uniqueness of the solution in the sens of \eqref{sol_strong_3}-\eqref{sol_strong_4}.
     The last step of the proof is the application of the Lyapunov analysis on the quotient space $\mathtt{H}/N(B)$ which has been show to be a Hilbert space in Lemma~\ref{lemma}.  
       
       The key point is using $B^2=B$ on the quotient space $\mathtt{H}/N(B)$ for system \eqref{sys_partial} and to consider the following candidate as Lyapunov function,
      \begin{align}
          V(t)=\frac12\scal{X(t)}{BX(t)}.
      \end{align}
For the right derivative of $V$ along a strong solution, one has 
  \begin{align}
  \frac{d^+}{dt} V(t) &\in  \{-\langle BX(t), (I-B)Y(t) - Bd(t) \notag\\&+ k_2 \lceil BX(t) \rfloor^\alpha \notag\\
  &\qquad + Bk Z(t) \rangle, Y(t)\in AX(t), Z(t)\in \lceil BX(t) \rfloor^0 \}\\
  \frac{d^+}{dt} V(t) &\in \{  -\langle X(t), B((I-B)Y(t) - d(t) \notag\\&+ k_2 \lceil BX(t) \rfloor^\alpha \notag\\
  &\qquad + k Z(t)) \rangle, Y(t)\in AX(t), Z(t)\in \lceil BX(t) \rfloor^0 \} \\
  \frac{d^+}{dt} V(t) &\in \{  -\langle BX(t), -d(t) + k_2 \lceil BX(t) \rfloor^\alpha\notag\\
  &\qquad + kZ(t)\rangle, Z(t)\in \lceil BX(t) \rfloor^0 \}.
  \end{align}
  It finally holds
  \begin{align}
      \frac{d^+}{dt}V(t)\leq& -k_2\norm{X(t)}^{\alpha+1}-(k-\bar d)\norm{X(t)}\\
      =&-k_2V(t)^{\frac{\alpha+1}{2}}-(k-\bar d)\sqrt{V(t)}.
  \end{align}
and one concludes by using Lemma~\ref{lemma1}. 
\end{proofof}
\begin{remark}
It is worth noting that the hypothesis of $(I-B)A$ monotone is not utilized in the convergence analysis, but it serves as a sufficient condition to establish well-posedness. It actually could be relax to $(I-B)A + c I $ is maximal monotone for some $c\in\mathbb{R}$, then $ (I-B)A  $ is the generator of a $ C^0 $-semigroup, as discussed in Chapter 4 of \cite{barbu1993analysis}. 
Noting that $ \mathcal{D}(BA) \subset \mathcal{D}(A) $, we apply Theorem~\ref{theo_rockafellar} to establish that the solution is well-posed.
  \end{remark}

  \begin{proofof}{Main Theorem~\ref{theo_partial_2}}
Let $A:\mathtt{H}\to 2^\mathtt{H}$ be a maximal monotone relation. Using Rockafellar's result Theorem~\ref{theo_rockafellar} and Lemma~\ref{lem_par}, we establish that the relation $x\to Ax + k\lceil B x \rfloor^0 + k_2\lceil B x \rfloor^\alpha$ is also maximal monotone. Let us consider the following Lyapunov functional
    \begin{align}
    V(t)=\frac12\scal{X(t)}{BX(t)}.
    \end{align}
  For the right derivative in time of $V$ along a strong solution, one has 
   
    \begin{align}
    \frac{d^+}{dt} V(t) &\in  \{-\langle BX(t), Y(t) - Bd(t) +k_2\lceil BX(t) \rfloor^\alpha  + BkZ(t)) \rangle, \notag\\
    &\qquad Y(t)\in AX(t), Z(t)\in \lceil BX(t) \rfloor^0 \},
    \end{align}
    Using now the condition \eqref{ABx}, it holds
    \begin{align}
        \frac{d^+}{dt} V(t)\in & \{-\scal{BX(t)}{Y(t)} -\langle BX(t), -Bd(t) \notag\\& +k_2 \lceil BX(t) \rfloor^\alpha \notag\\
        & + Bk Z(t) \rangle, Y(t)\in AX(t), Z(t)\in \lceil BX(t) \rfloor^0 \} \\
        \leq & \omega_1\norm{BX(t)}+\omega_2\norm{BX(t)}^{\alpha+1} \notag\\&-k_2\norm{BX(t)}^{\alpha+1}-(k-\bar d)\norm{BX(t)}\\
        = & -(k_2-\omega_2)V(t)^{\frac{\alpha+1}{2}}\notag\\&-(k-\omega_1-\bar d) \sqrt{V(t)}.
    \end{align}
    The finite-time convergence is thus established, this concludes the proof of the theorem.
    \end{proofof}

\begin{proofof}{Main Theorem~\ref{theo_inhomo_fixed}}
By setting $ B = I $, where $ I $ is the identity map, the conditions of Main Theorem~\ref{theo_partial_2} are satisfied with $ \omega_1 = 0 $ and $ \omega_2 = 0 $, thereby completing the proof.
\end{proofof}

\section{Exemples}
\subsection{Application to a Heat Equation with Modal Projection}
\label{application_heat_modal}

To illustrate the practical relevance of Main Theorem~\ref{theo_partial_2}, we consider a distributed parameter system where control actuation is restricted to a finite-dimensional subspace of the state space. This scenario corresponds to {modal control}, where only the dominant low-frequency modes are actively stabilized using discontinuous feedback, while the remaining high-frequency modes rely on the natural dissipation of the plant.

\subsubsection{Problem Statement}
Let $\Omega = (0,1)$ be the spatial domain. We consider the following controlled heat equation:
\begin{subnumcases}{\label{sys_heat_modal}}
    u_t(t,x) = u_{xx}(t,x) + U(t,x) + d(t,x), \quad t>0, \; x \in \Omega, \\
    u(t,0) = u(t,1) = 0, \\
    u(0,x) = u_0(x).
\end{subnumcases}
Here, $u(t,\cdot)$ represents the state profile, $d(t,\cdot)$ is a bounded external disturbance conform to the theorems assumption, and $U(t,\cdot)$ is the control input.

We define the Hilbert space $\mathtt{H} = L^2(0,1)$ endowed with the standard inner product $\scal{\cdot}{\cdot}$ and norm $\norm{\cdot}$. The system \eqref{sys_heat_modal} can be rewritten in the abstract form \eqref{sys_partial_2}:
\begin{align}
    \dot{X}(t) + A X(t) + k \lceil B X(t) \rfloor^0 + k_2 \lceil B X(t) \rfloor^\alpha \ni B d(t),
\end{align}
where $X(t) = u(t,\cdot)$. Let $A = -\partial_{xx}$ with domain $\mathcal{D}(A) = H^2(0,1) \cap H^1_0(0,1)$. It is well-known that $A$ is a positive, self-adjoint, unbounded operator with compact resolvent \cite{pazy2012semigroups}. Its spectrum consists of eigenvalues $\lambda_k = (k\pi)^2, \ k\in [1,\infty)$ associated with the eigenfunctions $\phi_k(x) = \sqrt{2}\sin(k\pi x)$, which form an orthonormal basis of $\mathtt{H}$. We define $B$ as the orthogonal projection onto the subspace spanned by the first $N$ eigenfunctions of $A$:
    \begin{align}\label{def_B_modal}
        B u = \sum_{k=1}^N \scal{u}{\phi_k} \phi_k, \quad \forall u \in \mathtt{H}.
    \end{align}
    By construction, $B$ is a bounded, linear, self-adjoint, and projective operator ($B^2 = B = B^*$).

\subsubsection{Verification of Assumptions for Main Theorem 3}
To apply Main Theorem~\ref{theo_partial_2}, we must verify the maximal monotonicity of $A$ and the coupling condition \eqref{ABx}.

Since $A$ is a positive self-adjoint operator, it is maximal monotone. The nonlinear control term $z \mapsto \lceil B z \rfloor^\alpha$ is maximal monotone by Lemma~\ref{lem_par}. Thus, the well-posedness conditions are satisfied. The structural relationship between $A$ and $B$, satisfy condition \eqref{ABx}. Since $B$ is defined via the spectral decomposition of $A$, the two operators commute on $\mathcal{D}(A)$.
\begin{lemma}\label{lemma_commute}
    For all $u \in \mathcal{D}(A)$, it holds $A B u = B A u$. Moreover, $\scal{A u}{B u} \geq \pi^2 \norm{B u}^2$.
\end{lemma}
\begin{proofof}{Lemma~\ref{lemma_commute}}
    Let $u \in \mathcal{D}(A)$. We can expand $u$ in the eigenbasis: $u = \sum_{k=1}^\infty c_k \phi_k$, where $\sum \lambda_k^2 |c_k|^2 < \infty$.
    Applying $B$ yields $B u = \sum_{k=1}^N c_k \phi_k$. Since this is a finite sum of smooth eigenfunctions, $B u \in \mathcal{D}(A)$.
    Then:
    \begin{align}
        A B u = A \left( \sum_{k=1}^N c_k \phi_k \right) = \sum_{k=1}^N c_k \lambda_k \phi_k.
    \end{align}
    Conversely:
    \begin{align}
        B A u = B \left( \sum_{k=1}^\infty c_k \lambda_k \phi_k \right) = \sum_{k=1}^N c_k \lambda_k \phi_k.
    \end{align}
    Thus $AB = BA$. For the inner product:
    \begin{align}
        \scal{A u}{B u} = \scal{A B u}{B u} = \sum_{k=1}^N \lambda_k |c_k|^2 \geq \lambda_1 \sum_{k=1}^N |c_k|^2 = \lambda_1 \norm{B u}^2.
    \end{align}
    Since $\lambda_1 = \pi^2 > 0$, the condition is satisfied.
\end{proofof}

Consequently, the coupling condition \eqref{ABx} required in Main Theorem~\ref{theo_partial_2} holds with $\omega_1 = 0$ and $\omega_2 = 0$:
\begin{align}
    \scal{y}{B x} \geq 0, \quad \forall y \in A x.
\end{align}
This implies that the uncontrolled dynamics does not destabilize the controlled subspace; instead, they contribute to dissipation.

\subsubsection{Control Law and Convergence Result}
We implement the following discontinuous feedback law acting only on the first $N$ modes:
\begin{align}\label{control_law_modal}
    U(t,x) = -k \lceil B u(t,\cdot) \rfloor^0 - k_2 \lceil B u(t,\cdot) \rfloor^\alpha,
\end{align}
with gains $k > \bar{d}$ and $k_2 > 0$.

\begin{theorem}\label{theo_modal_result}
    Consider system \eqref{sys_heat_modal} with control \eqref{control_law_modal}. If $k > \bar{d}$, then the projected state $B u(t)$ converges to zero in fixed time $T_{\max}$, independent of the initial condition $u_0$. Specifically:
    \begin{align}
        \norm{B u(t)} = 0, \quad \forall t \geq T_{\max} = \frac{1}{k - \bar{d}} + \frac{1}{k_2 (\alpha - 1)}.
    \end{align}
    Moreover, the residual state $(I-B)u(t)$ converges exponentially to zero as $t \to \infty$.
\end{theorem}
\begin{proof}
The proof is immediate after using Lemma~\ref{lemma_commute} and Theorem~\ref{theo_inhomo_fixed_partial_feed}.
\end{proof}

This result highlights the efficiency of partial stabilization. The controller expends energy only to suppress the dominant low-frequency modes (which are most visible and often most unstable in practice) in fixed time. The high-frequency residual modes, which are naturally damped by the Laplacian (since $\lambda_k \to \infty$), are left uncontrolled but still decay exponentially. This avoids the ``spillover effect'' often encountered in finite-dimensional controller designs for PDEs \cite{balogoun2025sliding}, as the infinite-dimensional stability is guaranteed by Theorem~\ref{theo_partial_2}.

\subsection{Heat with memory with linear coupling term\label{heat_lien}}
The following example illustrates a case where  fixed time is not guarantee, nevertheless using different development finite time convergence is achieved.

The result of Theorem~\ref{theo_partial_2} can be applied to the heat equation with a memory term. Partial differential equations with memory have a long and rich history, originating from the classical works of Maxwell \cite{Maxwell1867} and Boltzmann \cite{Boltzmann1874,Boltzmann1878}, and rigorously formalized by Volterra \cite{Volterra1912,Volterra1913}. These systems reflect the intrinsic memory of the process, where the present state depends on past evolution. They have been extensively studied in viscoelasticity, thermodynamics, and heat conduction (see, e.g., \cite{ColemanGurtin1967,Christensen1982,Cattaneo1958,GurtinPipkin1968,Amendola2012,Fabrizio2010}). Questions regarding well-posedness, asymptotic behavior, and controllability remain active areas of research \cite{ChavesSilva2017,Yong2011,Wang2022}.

Let us consider the following coupled system on the spatial domain $x \in (0,1)$:
\begin{subnumcases}{\label{sys_heat}}
v_{t}(t,x) = v_{xx}(t,x) - w(t,x) + U(v(t,\cdot),x) + d(t,x),  \\
v(t,0) = v(t,1) = 0, \\
w_t(t,x) = -\beta w(t,x) + \eta v(t,x),\\
v(0,\cdot) = v_0, \quad w(0,\cdot) = w_0,
\end{subnumcases}
where $\beta > 0$ is the memory decay rate, $\eta > 0$ is the coupling coefficient, and $d(t,\cdot)$ is a bounded disturbance such that $\forall t\geq 0, \ \|d(t,\cdot)\|_{L^2} \leq \bar{d}$.
Using the variation of constants formula, the memory term can be expressed as:
\begin{align}
    w(t,x)=e^{-\beta t}w_0(x)+\eta\int_0^te^{-\beta (t-s)}v(s,x)ds.
\end{align}
The control objective is to stabilize the temperature profile $v(t,\cdot)$ to zero in finite time, despite the presence of the memory term $w$ and the disturbance $d$.

We define the abstract state $X(t)=[v(t,\cdot),\ w(t,\cdot)]^\top \in \mathtt{H} = L^2(0,1) \times L^2(0,1)$.
We endow $\mathtt{H}$ with the weighted scalar product:
\begin{align}
    \scal{Z}{Q}_{\mathtt{H}}=\int_0^1 \left(\eta v_z v_q + w_z w_q\right) dx, \quad \forall Z=[v_z, w_z]^\top, Q=[v_q, w_q]^\top.
\end{align}
This choice of norm corresponds to the total physical energy of the system.
The abstract formulation is 
\begin{align}
    Az=\begin{bmatrix}
    -z_1 ''+z_2 \\ +\beta z_2 -\eta z_1
    \end{bmatrix}, \ z\in\mathcal{D}(A)=\{ z\in H^2\times L^2,\ z_1(0)=z_1(1)=0\},
\end{align}
and 
\begin{align}
    B=\begin{bmatrix}
    1&0\\ 0&0
    \end{bmatrix}.
\end{align}
It is direct that $B$ is a auto-adjoint projection, however the coupling condition \eqref{ABx} is not satisfy, in other words we cannot control $w$ by the $B$ action in this case. Nevertheless, using the total energy dissipation which ensures that the memory term $w(t)$ decays exponentially, we can show that the sliding mode control can dominate the dynamics of $w(t)$ after a transient period.

\begin{theorem}\label{theo_ex_1}
Consider system \eqref{sys_heat} with the control law 
\begin{align}\label{def_U1}
    U(v(t,\cdot),x) = -k\lceil v(t,\cdot) \rfloor^0 - k_2\lceil v(t,\cdot) \rfloor^\alpha,
\end{align}
where $k > \bar{d}$, $k_2 \geq 0$, and $\alpha \geq 0$. 
Then, there exists a finite positive time $T$ such that
\begin{align}
    \Vert v(t,\cdot) \Vert = 0,\quad \forall t\geq T.
\end{align}
\end{theorem}

\begin{proofof}{Theorem \ref{theo_ex_1}}
The existence of solution is given by the premise of Theorem~\ref{theo_inhomo_fixed_partial_feed}. The convergence proof proceeds in two steps: firstly, we establish the exponential decay of the total energy to bound the memory term $w(t)$; secondly, we prove the finite-time convergence of $v(t)$ once the memory term becomes sufficiently small. 

Consider the total energy Lyapunov functional candidate:
\begin{align}
    E(t) = \frac{\eta}{2} \norm{v(t,\cdot)}^2 + \frac{1}{2} \norm{w(t,\cdot)}^2.
\end{align}
Computing the time derivative of $E(t)$ along the trajectories of \eqref{sys_heat}:
\begin{align}
    \frac{d^+}{dt} E(t) &= \eta \langle v, v_t \rangle + \langle w, w_t \rangle \notag \\
    &\in \{ \eta \langle v, v_{xx} - w + Y + d \rangle + \langle w, -\beta w + \eta v \rangle, Y\in U\}.
\end{align}
Expanding the terms and using the boundary conditions $v(t,0)=v(t,1)=0$:
\begin{align}
    \frac{d^+}{dt} E(t) &\in \{ \underbrace{\eta \langle v, v_{xx} \rangle}_{-\eta \|v_x\|^2} \underbrace{-\eta \langle v, w \rangle + \eta \langle w, v \rangle}_{= 0} \notag \\
    &\quad - \beta \norm{w}^2 + \eta \langle v, Y \rangle + \eta \langle v, d \rangle, Y\in U\}.
\end{align}
The cross-terms cancel and by substituting the control law \eqref{def_U1} and using the fact that $\langle v, \lceil v \rfloor^0 \rangle = \|v\|$, the set of right derivatives  reduces to a element:
\begin{align}
    \frac{d^+}{dt} E(t) &= -\eta \|v_x\|^2 - \beta \norm{w}^2 - \eta k \norm{v} - \eta k_2 \norm{v}^{\alpha+1} + \eta \langle v, d \rangle \notag \\
    &\leq -\eta \|v_x\|^2 - \beta \norm{w}^2 - \eta (k - \bar{d}) \norm{v}.
\end{align}
Since $k > \bar{d}$, the term $-\eta (k - \bar{d}) \norm{v}$ is non positive. Dropping this term provides a conservative upper bound:
\begin{align}
    \frac{d^+}{dt} E(t) \leq -\eta \|v_x\|^2 - \beta \norm{w}^2.
\end{align}
Using Poincaré's inequality $\|v_x\|^2 \geq \pi^2 \|v\|^2$, we obtain:
\begin{align}
    \frac{d^+}{dt} E(t) \leq -\eta \pi^2 \|v\|^2 - \beta \norm{w}^2.
\end{align}
Recalling the definition of $E$, it holds
\begin{align}
    \frac{d^+}{dt} E(t) \leq -2\pi^2 \left(\frac{\eta}{2} \|v\|^2\right) - 2\beta \left(\frac{1}{2} \norm{w}^2\right) \leq -\lambda E(t),
\end{align}
where $\lambda = \min(2\pi^2, 2\beta) > 0$. This implies exponential decay of $t\mapsto E(t)$ to zero and hence exponential decay of the memory term satisfies $t\mapsto\|w(t,\cdot)\|$ to zero as well.

Consider now the Lyapunov function for the state $v$, i.e.,  $V_v(t) = \frac{1}{2} \norm{v(t,\cdot)}^2$. Its right time derivative satisfies
\[\frac{d^+}{dt} V_v(t) \in \{ \langle v, v_{xx} - w + Y + d \rangle ,Y\in U\},
\]
implying that
\begin{align}
\frac{d^+}{dt} V_v(t) &\leq - \langle v, w \rangle - k \|v\| + \bar{d} \|v\| \notag \\
&\leq -\pi^2 \|v\|^2 - (k - \bar{d} - \|w(t)\|) \|v\|.\label{eq:derV}
\end{align}

Since $\|w(t,\cdot)\| \to 0$ exponentially, there exists a time $T_1$ such that for every $t \geq T_1$, one has
\begin{equation}\label{eq:bound-w}
    \|w(t)\| \leq \frac{k - \bar{d}}{2}.
\end{equation}
For $t \geq T_1$, \eqref{eq:derV} and \eqref{eq:bound-w} yield that
\begin{align}
    \frac{d^+}{dt} V(t) \leq -\frac{k - \bar{d}}{2} \|v\| = -\frac{k - \bar{d}}{\sqrt{2}} \sqrt{V(t)}.
\end{align}
That differential inequality guarantees that $V(t)$ reaches zero in a finite time $T_2$ after $T_1$. Thus, $v(t)$ converges to zero in finite time $T = T_1 + T_2$.
\end{proofof}

\subsection{Heat with convex potential coupling term\label{heat_convexe}}

Let us consider the system
defined by
\begin{subnumcases}{\label{sys_heat_2}}
v_{t}(t,x) = v_{xx}(t,x) - \partial_v \Phi(v(t,x),w(t,x)) + U(v(t,\cdot),x) + d(t),  \\
v(t,0) = v(t,1) = 0, \\
w_t(t,x) = -\beta w(t,x) + \partial_w \Phi(v(t,x),w(t,x)),\\
v(0,\cdot) = v_0, \quad w(0,\cdot) = w_0,
\end{subnumcases}
with $x\in [0,1]$, $t\geq 0$ and $\Phi : \mathbb{R}^2 \to \mathbb{R}$ is a coupling potential. We have the following result.
\begin{theorem}\label{theo_coupling}
If the coupling potential $\Phi$ is convex verifying
\begin{align}\label{eq:condPhi}
    v\partial_v \Phi (v,w)\geq -k_1|v|,
    \quad (v,w)\in\mathbb{R}^2,
\end{align}
for some positive constant $k_1$,
then \eqref{sys_heat_2} satisfies the conditions of Theorem~\ref{theo_partial_2}, with
\begin{align}\label{def_U2}
    U(v(t,\cdot))=-k\lceil v(t,\cdot) \rfloor^0-k_2\lceil v(t,\cdot) \rfloor^\alpha.
\end{align} Therefore, for $k > \bar{d} + k_1$, $k_2 > 0$, and $\alpha > 1$, there exists $T_0>0$ such that
\begin{align}
    \Vert v(t,\cdot)\Vert = 0 ,\  \forall t\geq T_0.
\end{align}
\end{theorem}

\begin{remark}
Any convex potential with bounded partial derivative in $v$ satisfies \eqref{eq:condPhi}.
\end{remark}

\begin{proof}
With the state as $X(t)=[v(t,\cdot),\ w(t,\cdot)]$, the operator $B$ is defined as
\begin{align}\label{def_B}
B=\begin{bmatrix}
    1&0\\0&0
\end{bmatrix}\in L^2\times L^2\to L^2\times L^2.
\end{align}

The operator $A$ is now defined to include the nonlinear coupling terms:
\begin{align}\label{def_A}
    \forall z\in \mathcal{D}(A), Az=\begin{bmatrix}
        -z_1'' - \partial_v \Phi(z_1, z_2) \\ \beta z_2 + \partial_w \Phi(z_1, z_2)
    \end{bmatrix},
\end{align}
with $\mathcal{D}(A)=\{H^2\times L^2, z_1(0)=0, z_1(1)=0\}$.
Considering the Hilbert space $L^2\times L^2$ with the associated scalar product
\begin{align}
    \scal{z}{q}=\int_0^1 (\eta z_1q_1+z_2q_2) dx.
\end{align}

We verify that the operator $A$ defined in \eqref{def_A} satisfies the condition \eqref{ABx}.
For any $z \in \mathcal{D}(A)$, we have:
\begin{align}
    \scal{Az}{Bz} &= \scal{\begin{bmatrix} -z_1'' - \partial_v \Phi \\ \beta z_2 + \partial_w \Phi \end{bmatrix}}{\begin{bmatrix} z_1 \\ 0 \end{bmatrix}} \notag \\
    &= \int_0^1 \eta z_1 (-z_1'' - \partial_v \Phi) \, dx \notag \\
    &= \int_0^1 \eta z_1'^2 \, dx - \eta \int_0^1 z_1 \partial_v \Phi \, dx.
\end{align}
Since the first term is non-negative, we have:
\begin{align}
    \scal{Az}{Bz} &\ge - \eta \int_0^1 z_1 \partial_v \Phi \, dx.
\end{align}
Using the property of the convex potential $\Phi$, we know that $z_1 \partial_v \Phi \ge -k_1 |z_1|$. Thus:
\begin{align}
    \scal{Az}{Bz} &\ge -\eta k_1 \norm{z_1}.
\end{align}
This matches condition \eqref{ABx} with $\omega_1 = 0$ (no nonlinear term in the coupling bound) and $\omega_2 = \eta k_1$ (linear bound).
Note that $A$ is maximal monotone as it is the sum of a maximal monotone linear operator (Laplacian + damping) and a gradient of a convex function.

\end{proof}

\section{Numerical Applications}
\subsection{Wave Equation with Modal Partial Actuation: Finite-Time vs. Fixed-Time Stabilization}

\begin{figure}
    \centering
    \includegraphics[width=\linewidth]{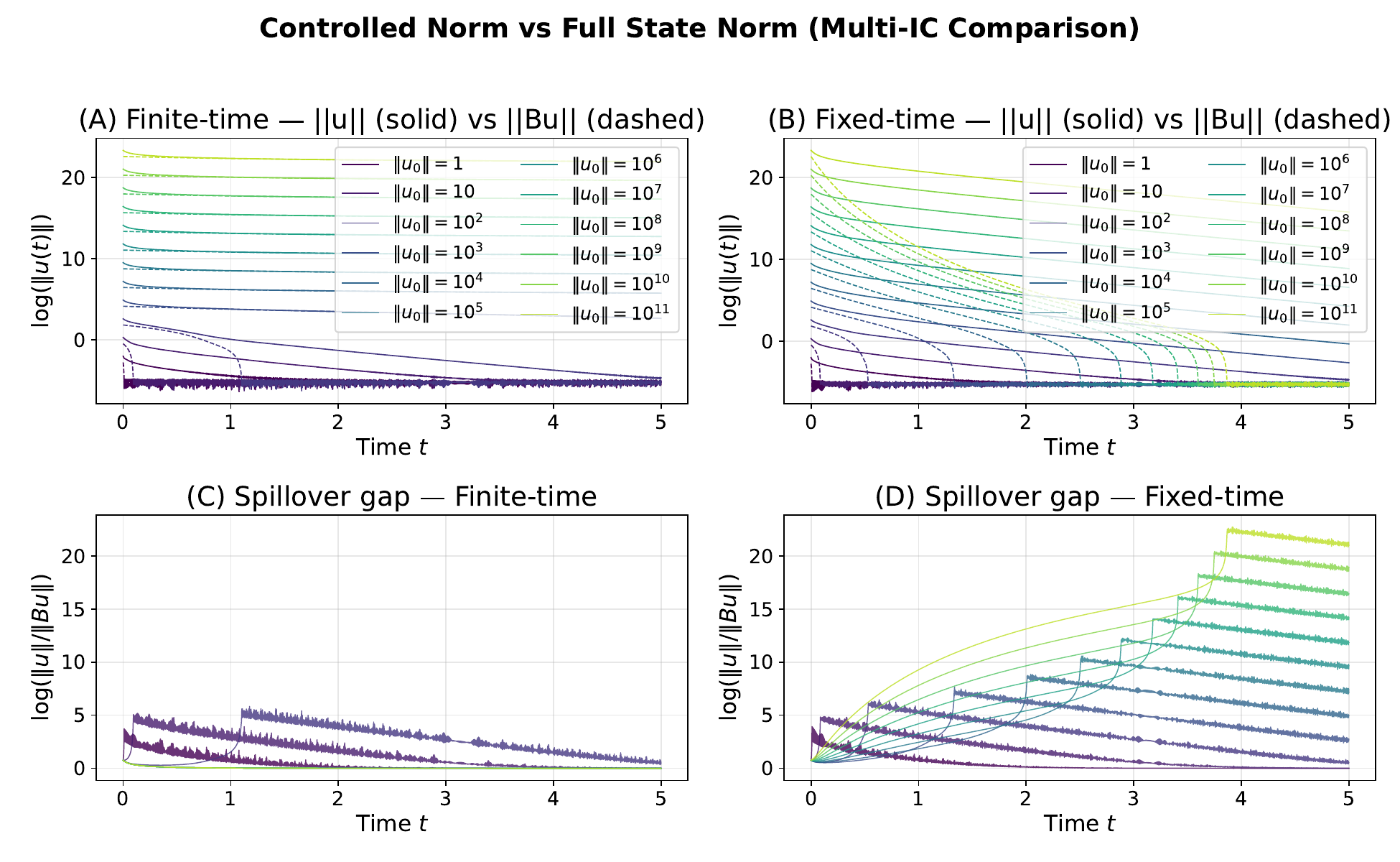}
    \caption{Comparison between finite-time stabilization ($k=3,k_2=0$) and fixed-time stabilization ($k=3,k_2=20/9,\alpha=1.1$).}
    \label{fig:blabla}
\end{figure}
\begin{figure}
    \centering
    \includegraphics[width=\linewidth]{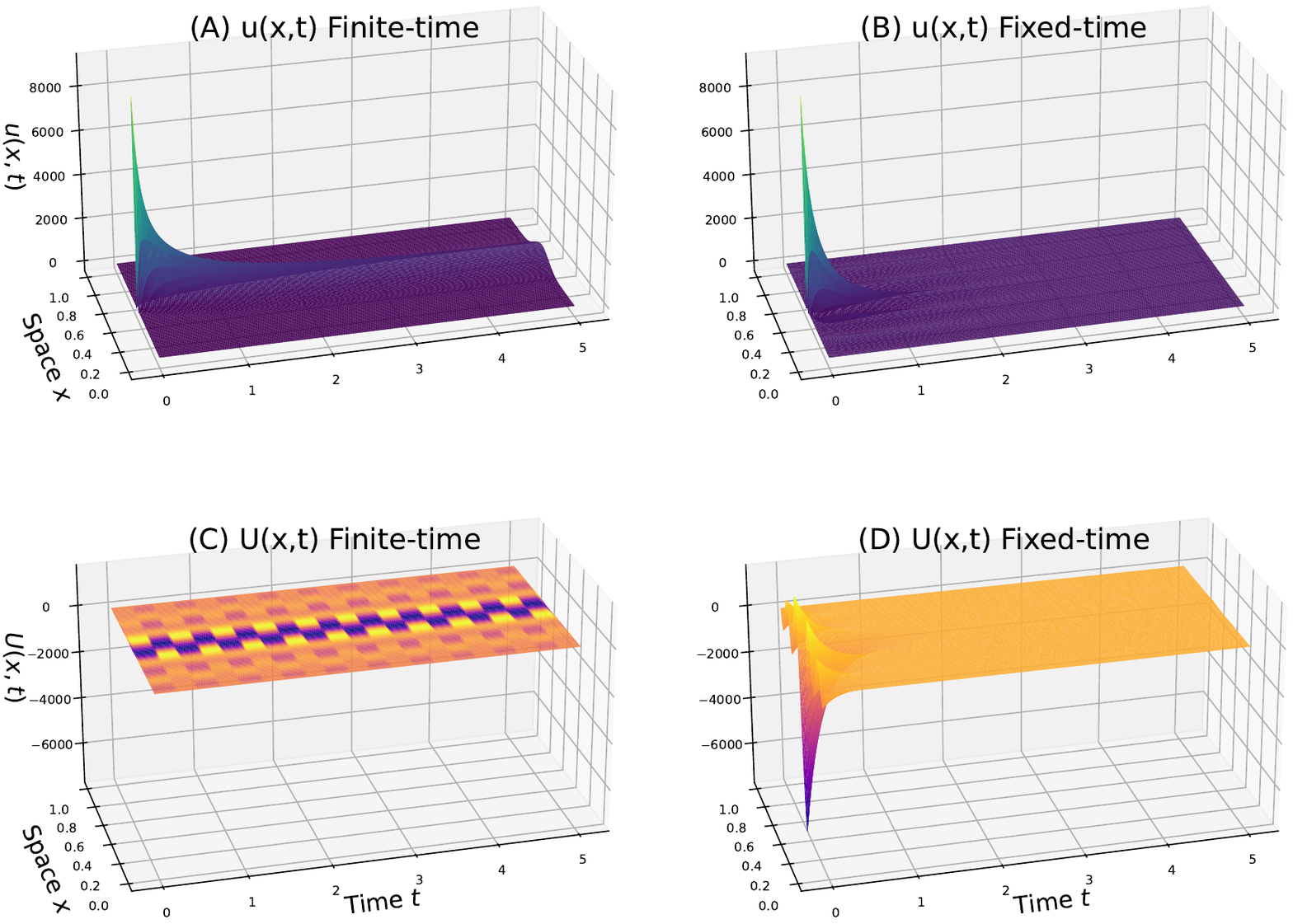}
    \caption{Comparison between finite-time stabilization ($k=3,k_2=0$) and fixed-time stabilization ($k=3,k_2=20/9,\alpha=1.1$).}
    \label{fig:blabla2}
\end{figure}

The controlled heat equation $u_t = \nu u_{xx} + U + d$ with homogeneous Dirichlet boundary conditions on the spatial domain $(0,1)$ is solved via modal decomposition. The solution is expanded as $u(t,x) = \sum_{k=1}^M c_k(t)\phi_k(x)$ using the eigenbasis $\phi_k(x)=\sqrt{2}\sin(k\pi x)$ of the Dirichlet Laplacian, yielding a diagonal system of ODEs:
\begin{equation}
    \dot{c}_k + \lambda_k c_k = U_k + D_k, \quad \lambda_k = \nu(k\pi)^2, \quad k=1,\dots,M.
\end{equation}
Time integration employs a Crank–Nicolson scheme for the linear diffusive part, while the nonlinear control input is treated explicitly. The simulation retains $M=100$ spectral modes; however, the control operator $B$ acts exclusively on the first $N_c=10$ modes. The remaining $90$ higher-frequency modes evolve in open loop under the sole action of diffusion and external disturbance, thereby introducing a spillover phenomenon. However as said in the Theorem~\ref{theo_modal_result}, the spillover does not destabilized the system.

Both control laws share the discontinuous structure $U = -k\lceil Bc\rfloor^0 - k_2\lceil Bc\rfloor^\alpha$. The finite-time (FT) configuration sets $k_2=0$, retaining only the term $-k\lceil Bc\rfloor^0$; its convergence time scales with the initial condition magnitude. The fixed-time (FxT) configuration activates the nonlinear term with $k_2=20/9$ and $\alpha=1.1$, guaranteeing a uniform convergence bound $T_{\max}=1/(k-\bar{d}) + 1/k_2(\alpha-1)=5$, independent of the initial state. In both cases, the common gain $k=3$ satisfies the reaching condition $k > \bar{d}=1$. A time-varying disturbance $d(t,x) = B\sin(t)$, uniform in space, is applied and projected onto the eigenbasis at each time step.

Figure~\ref{fig:blabla} overlays two metrics per simulation: $\log\|Bu(t)\|$ (controlled modal norm, dashed lines) and $\log\|u(t)\|$ (full state norm, solid lines), for $12$ initial condition amplitudes ranging from $\|u_0\|=1$ to $10^{11}$. Panels A and B correspond to the FT and FxT laws, respectively; the vertical gap between each solid and dashed pair of identical color quantifies the energy spilled into the uncontrolled modes. Panels C and D plot this spillover ratio $\log(\|u\|/\|Bu\|)$ directly. A monotonically growing gap indicates a persistent leakage of energy into the uncontrolled subspace (observed under FT control), whereas a gap that saturates or decays indicates that the control action effectively bounds or recaptures the spillover (observed under FxT control). The bottom-row panels share a common $y$-axis range to facilitate direct visual comparison. 

Figure~\ref{fig:blabla2} depicts the full spatiotemporal evolution for a single large initial condition ($\|u_0\| = 10^4$, localized spike profile). Panels A and B show the physical state $u(x,t)$ under FT and FxT control, respectively. Panels C and D display the corresponding control input $U(x,t)$, revealing the spatial distribution and magnitude of the control effort. The control surfaces expose the distinct switching structures: the finite time law exhibits sustained high-frequency chattering across the entire spatial domain for the duration of the simulation, whereas the fixed time law generates an intense initial burst that rapidly attenuates once the state reaches the sliding surface. This spatial visualization complements the norm-based analysis by illustrating where the control effort concentrates and how the transient spatial structure differs between the two laws.

\subsection{Heat Equation with Convex Potential Coupling}

\begin{figure}
    \centering
    \includegraphics[width=\linewidth]{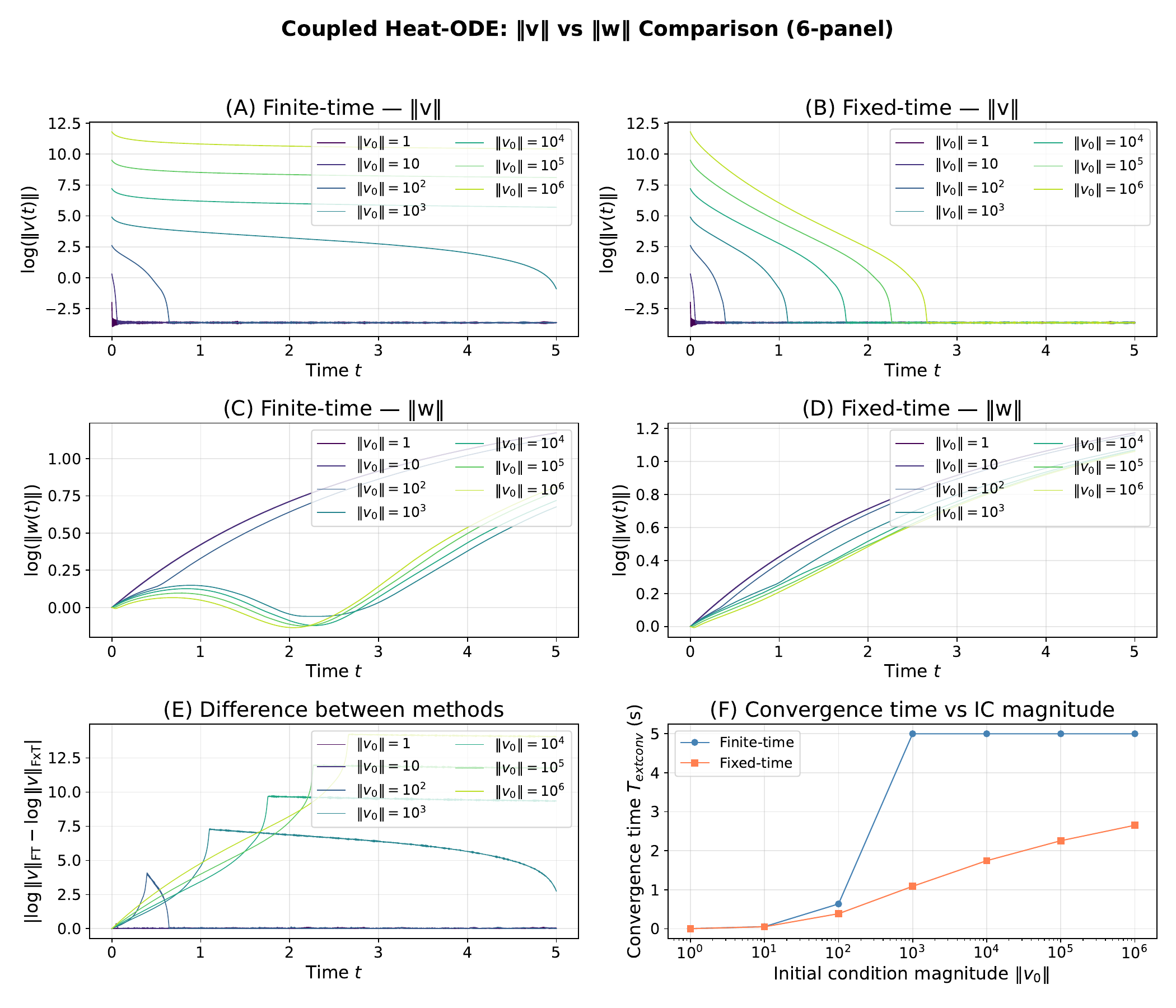}
    \caption{Comparison between finite-time stabilization ($k=3,k_2=0$) and fixed-time stabilization ($k=3,k_2=20/9,\alpha=1.1$).}
    \label{fig:blabla3}
\end{figure}
\begin{figure}
    \centering
    \includegraphics[width=\linewidth]{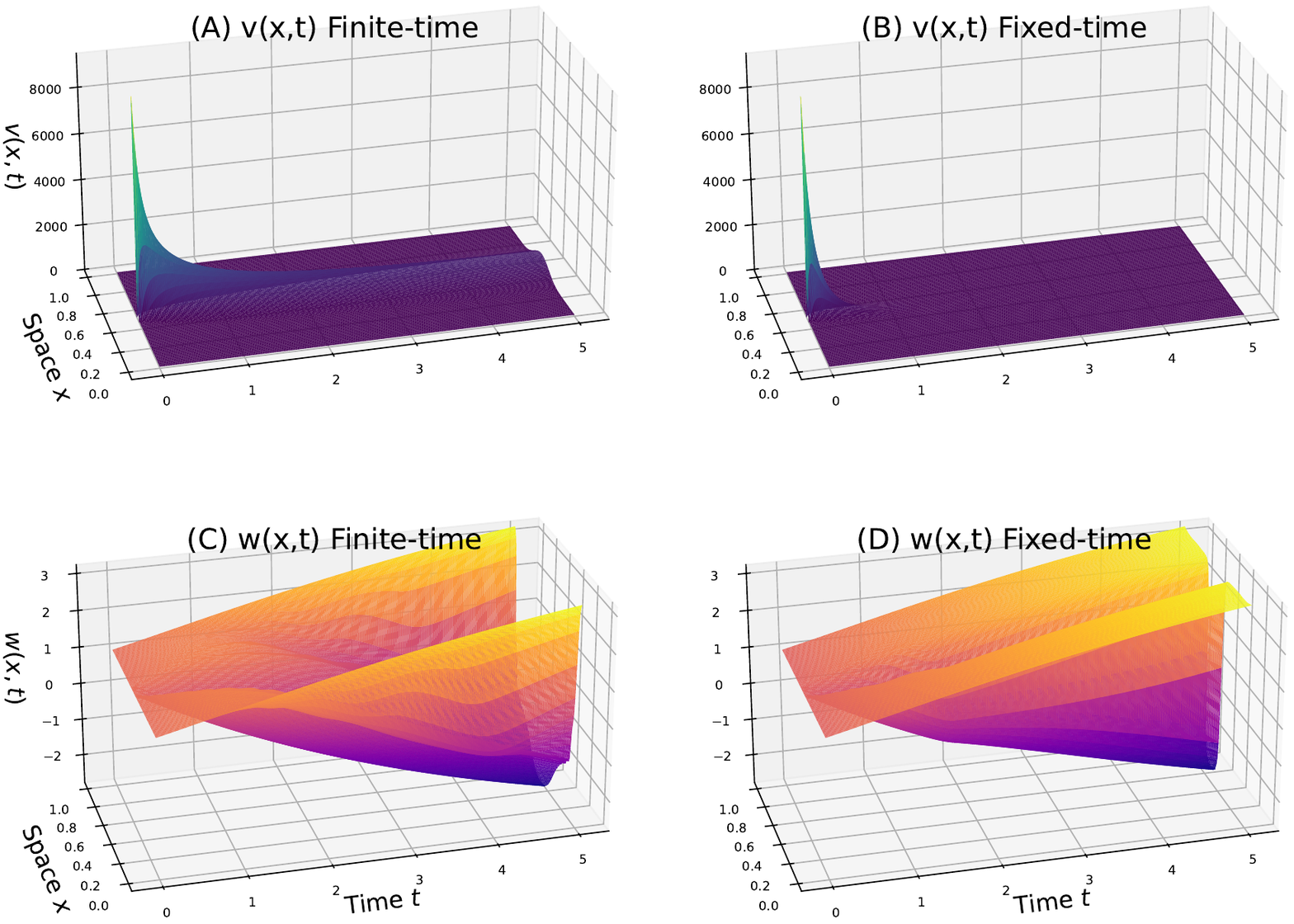}
    \caption{Comparison between finite-time stabilization ($k=3,k_2=0$) and fixed-time stabilization ($k=3,k_2=20/9,\alpha=1.1$).}
    \label{fig:blabla4}
\end{figure}

The second benchmark considers a coupled PDE-ODE system defined on $(0,1)$ with Dirichlet boundary conditions for the PDE component:
\begin{subequations}
\begin{align}
    v_t &= v_{xx} - \partial_v\Phi(v,w) + U(v) + d(t), \\
    w_t &= -\beta w + \partial_w\Phi(v,w).
\end{align}
\end{subequations}
The coupling potential $\Phi(v,w) = k_1\sqrt{1+(v-w)^2}$ has globally bounded derivatives ($|\partial_v\Phi|, |\partial_w\Phi| \le k_1$), such that the coupling acts as an additional matched disturbance of known bound $k_1$ on the $v$-equation. The control law is $U(v) = -k\lceil v\rfloor^0 - k_2\lceil v\rfloor^\alpha$. The $v$-equation is discretized via modal decomposition using $N=100$ spectral modes. The $w$-equation, being a continuum of uncoupled ODEs parameterized by $x$, is integrated using a Crank–Nicolson scheme on a uniform physical grid. Key parameters are: coupling strength $k_1=1.0$, damping $\beta=0.2$, sliding gain $k=5.0$ (satisfying $k > \bar{d}+k_1 = 2$), FxT gain $k_2=20/9$, exponent $\alpha=1.1$, and thermal diffusivity $\nu=0.1$ (individual cases) or $\nu=0.001$ (multi-initial-condition comparison). Under these parameters, the theoretical fixed-time convergence bound is $T_{\max} = 1/(k-\bar{d}-k_1) + 1/(k_2(\alpha-1)) \approx 4.83$~s.

Figure~\ref{fig:blabla3} presents a comprehensive convergence analysis for both state components across seven initial condition magnitudes ($\|v_0\| = 1, 10, 10^2, \dots, 10^6$) at $\nu=0.001$ with $w_0=1$. Panels A and B show $\log\|v(t)\|$ for FT and FxT stabilization, respectively; Panels C and D show the corresponding $\log\|w(t)\|$; Panel E plots the absolute difference $\bigl|\log\|v\|_{\text{FT}} - \log\|v\|_{\text{FxT}}\bigr|$; and Panel F quantifies the convergence time $T_{\text{conv}}$ (defined as the time to reach $\log\|v\| < -3$) as a function of $\|v_0\|$ on a log-linear scale. The primary observation is that $\|v(t)\|$ converges to zero in finite time under both schemes, but the FxT convergence time is bounded invariantly to $\|v_0\|$, whereas the FT convergence time shifts right with increasing initial energy. The $w$-dynamics (Panels C–D) exhibit a distinct two-stage behavior: an initial transient where $w$ evolves under the influence of the coupling term $\partial_w\Phi$, followed by a pure exponential decay at rate $\beta$ once $v$ reaches zero and $\partial_w\Phi = -\partial_v\Phi$ vanishes. 

Figure~\ref{fig:blabla4} illustrates the spatiotemporal surfaces $v(x,t)$ and $w(x,t)$ for a large spike initial condition ($\|v_0\| = 10^4$). The $v$-surfaces (Panels A, B) show the expected collapse toward zero within the first $2$–$3$ s under both laws, with the FxT law exhibiting a marginally steeper initial descent. The $w$-surfaces (Panels C, D) reveal the coupling dynamics: starting from a uniform profile $w_0=3$, the $w$-field develops spatial structure mirroring $-v$ through $\partial_w\Phi$, then relaxes toward zero (in a longer horizon) as $v$ is stabilized. The FxT $w$-panel (D) shows a faster return to zero compared to the FT panel (C), consistent with the earlier stabilization of $v$ removing the coupling forcing term sooner. The bottom-row panels include explicit time and space axis labels, while the top-row panels omit them for visual clarity, adhering to a consistent mosaic layout convention.
 
\section{Conclusions and future work}
In this paper, we study the finite-time stabilization of a class of abstract evolution systems and extend our results to fixed-time stabilization, where the settling time can be uniformly bounded independently of the initial conditions. Both analyses consider the presence of matched $L_\infty$ perturbations. For finite-time stabilization, we show existence of strong solutions within the framework of the abstract Cauchy problem, while for fixed-time stabilization, we employ a Lyapunov-based approach inspired by methods developed for finite-dimensional systems, extended to functional spaces endowed with appropriate norms. Using the theory of maximal monotone operators, we establish the well-posedness of the closed-loop system in both cases. A theoretical and numerical example based on the heat equation with a memory term illustrates the applicability of our results and highlights the gap between first-order and second-order sliding modes.

As a next step, we consider the second-order sliding-mode (twisting algorithm) problem
\begin{align*}
    v_{tt} -v_{xx}+ d(t)\in  - k_1\sign{v_t} - k_2\sign{v}.
\end{align*}
To the best of our knowledge, this problem remains open on the convergence and well-posedness properties since the regularization techniques used by \cite{orlov2020nonsmooth} and the viability method \cite{levaggi2002infinite} do not provide strong solutions.

\bibliographystyle{abbrv}
\bibliography{sn-bibliography,applications_heatwithmemory}

@article{Maxwell1867,
  author = {J. Maxwell},
  title = {On the dynamical theory of gases},
  journal = {Phil. Trans. Roy. Soc. London},
  volume = {157},
  pages = {49--88},
  year = {1867}
}

@article{Boltzmann1874,
  author = {L. Boltzmann},
  title = {Zur theorie der elastischen nachwirkung},
  journal = {Wien. Ber.},
  volume = {70},
  pages = {275--306},
  year = {1874}
}

@article{Boltzmann1878,
  author = {L. Boltzmann},
  title = {Zur theorie der elastischen nachwirkung},
  journal = {Wien. Ber.},
  volume = {5},
  pages = {430--432},
  year = {1878}
}

@article{Volterra1912,
  author = {V. Volterra},
  title = {Sur les équations intégrro-différentielles et leurs applications},
  journal = {Acta Math.},
  volume = {35},
  pages = {295--356},
  year = {1912}
}

@book{Volterra1913,
  author = {V. Volterra},
  title = {Leçons sur les fonctions de lignes},
  publisher = {Gauthier-Villars, Paris},
  year = {1913}
}

@book{Amendola2012,
  author = {G. Amendola and M. Fabrizio and J. Golden},
  title = {Thermodynamics of materials with memory: Theory and applications},
  publisher = {Springer, New York},
  year = {2012}
}

@article{Cattaneo1958,
  author = {C. Cattaneo},
  title = {A form of heat conduction equation which eliminates the paradox of instantaneous propagation},
  journal = {C. R. Acad. Sci. Paris},
  volume = {247},
  pages = {431--433},
  year = {1958}
}

@article{ChavesSilva2017,
  author = {F. Chaves-Silva and X. Zhang and E. Zuazua},
  title = {Controllability of evolution equations with memory},
  journal = {SIAM J. Control Optim.},
  volume = {55},
  pages = {2437--2459},
  year = {2017}
}

@book{Christensen1982,
  author = {R. Christensen},
  title = {Theory of viscoelasticity, an introduction},
  publisher = {Academic Press, New York},
  year = {1982}
}

@article{ColemanGurtin1967,
  author = {B. Coleman and M. Gurtin},
  title = {Equipresence and constitutive equations for rigid heat conductors},
  journal = {Z. Angew. Math. Phys.},
  volume = {18},
  pages = {199--208},
  year = {1967}
}

@article{Fabrizio2010,
  author = {M. Fabrizio and C. Giorgi and V. Pata},
  title = {A new approach to equations with memory},
  journal = {Arch. Ration. Mech. Anal.},
  volume = {198},
  pages = {189--232},
  year = {2010}
}

@article{GurtinPipkin1968,
  author = {M. Gurtin and A. Pipkin},
  title = {A general theory of heat conduction with finite wave speeds},
  journal = {Arch. Ration. Mech. Anal.},
  volume = {31},
  pages = {113--126},
  year = {1968}
}

@article{Wang2022,
  author = {G. Wang and Y. Zhang and E. Zuazua},
  title = {Decomposition for the flow of the heat equation with memory},
  journal = {J. Math. Pures Appl.},
  volume = {158},
  pages = {183--215},
  year = {2022}
}

@article{Yong2011,
  author = {J. Yong and X. Zhang},
  title = {Heat equations with memory in anisotropic and non-homogeneous media},
  journal = {Acta Math. Sin. Engl. Ser.},
  volume = {27},
  pages = {219--254},
  year = {2011}
}

@ARTICLE{10742482,
  author={Polyakov, Andrey and Orlov, Yury},
  journal={IEEE Transactions on Automatic Control}, 
  title={Finite/Fixed-Time Homogeneous Stabilization of Infinite Dimensional Systems}, 
  year={2025},
  volume={70},
  number={4},
  pages={2560-2567},
  doi={10.1109/TAC.2024.3490989}}

@article{crandall1971generation,
  title={Generation of semi-groups of nonlinear transformations on general Banach spaces},
  author={Crandall, Michael G and Liggett, Thomas M},
  journal={American Journal of Mathematics},
  volume={93},
  number={2},
  pages={265--298},
  year={1971},
  publisher={JSTOR}
}

@article{vanspranghe2021velocity,
  title={Velocity stabilization of a wave equation with a nonlinear dynamic boundary condition},
  author={Vanspranghe, Nicolas and Ferrante, Francesco and Prieur, Christophe},
  journal={IEEE Transactions on Automatic Control},
  volume={67},
  number={12},
  pages={6786--6793},
  year={2021},
  publisher={IEEE}
}

@article{hayat2019quadratic,
  title={A quadratic {L}yapunov function for {S}aint-{V}enant equations with arbitrary friction and space-varying slope},
  author={Hayat, Amaury and Shang, Peipei},
  journal={Automatica},
  volume={100},
  pages={52--60},
  year={2019},
  publisher={Elsevier}
}

@article{pisano2012boundary,
  title={Boundary second-order sliding-mode control of an uncertain heat process with unbounded matched perturbation},
  author={Pisano, Alessandro and Orlov, Yury},
  journal={Automatica},
  volume={48},
  number={8},
  pages={1768--1775},
  year={2012},
  publisher={Elsevier}
}

@inproceedings{orlov2011boundary,
  title={Boundary second-order sliding-mode control of an uncertain heat process with spatially varying diffusivity},
  author={Orlov, Yury and Pisano, Alessandro and Usai, Elio},
  booktitle={2011 50th IEEE Conference on Decision and Control and European Control Conference},
  pages={1323--1328},
  year={2011},
  organization={IEEE}
}

@book{utkin2013sliding,
  title={Sliding modes in control and optimization},
  author={Utkin, Vadim I},
  year={2013},
  publisher={Springer Science \& Business Media}
}

@article{Levaggi2002sliding,
  author  = {Levaggi, L.},
  title   = {Sliding modes in Banach spaces},
  journal = {Differential and Integral Equations},
  volume  = {15},
  year    = {2002},
  pages   = {167--189}
}

@article{takahashi1976convergence,
  title={Convergence of difference approximation of nonlinear evolution equations and generation of semigroups},
  author={Takahashi, Tadayasu},
  journal={Journal of the Mathematical Society of Japan},
  volume={28},
  number={1},
  pages={96--113},
  year={1976},
  publisher={The Mathematical Society of Japan}
}

@inproceedings{benilan1972solutions,
  title={Solutions faibles d'{\'e}quations d'{\'e}volution dans les espaces de Hilbert},
  author={Benilan, Philippe and Brezis, Haim},
  booktitle={Annales de l'institut Fourier},
  volume={22},
  number={2},
  pages={311--329},
  year={1972}
}

@inproceedings{labbadi2025discretization,
  title={On the discretization of the implicit {L}yapunov function-based control},
  author={Labbadi, Moussa and Efimov, Denis},
  booktitle={CDC 2025-64th IEEE Conference on Decision and Control},
  year={2025}
}

@article{polyakov2015finite,
  title={Finite-time and fixed-time stabilization: Implicit {L}yapunov function approach},
  author={Polyakov, Andrey and Efimov, Denis and Perruquetti, Wilfrid},
  journal={Automatica},
  volume={51},
  pages={332--340},
  year={2015},
  publisher={Elsevier}
}

@book{Polyakov2025,
  author    = {Andrey Polyakov},
  title     = {Generalized Homogeneity in Systems and Control: Finite- and Infinite-Dimensional Systems},
  year      = {2025},
  publisher = {Springer Cham},
  edition   = {2},
  pages     = {XXIX, 950},
  isbn      = {978-3-031-79113-0},  
  url       = {https://link.springer.com/book/9783031791130}
}

@article{Orlov2025SlidingModeBoundaryControl,
  title        = {Sliding Mode Boundary Control of an Uncertain Parabolic PDE–ODE Cascade with Dynamic Actuation and Unmatched Disturbance},
  author       = {Yury Orlov and Alessandro Pilloni and Alessandro Pisano and Elio Usai},
  journal      = {IEEE Transactions on Automatic Control},
  year         = {2025},
  volume       = {PP},
  number       = {99},
  pages        = {1--8},
  month        = {Jan},
  doi          = {10.1109/TAC.2025.3648248},
}

@article{chitour2021one,
  title={One-dimensional wave equation with set-valued boundary damping: well-posedness, asymptotic stability, and decay rates},
  author={Chitour, Yacine and Marx, Swann and Mazanti, Guilherme},
  journal={ESAIM: Control, Optimisation and Calculus of Variations},
  volume={27},
  pages={84},
  year={2021},
  publisher={EDP Sciences}
}

@book{barbu1993analysis,
  title={Analysis and Control of Nonlinear Infinite Dimensional System},
  author={V. Barbu},
  publisher={Academic Press},
  address={Boston},
  year={1993}
}

@techreport{aubin1982differential,
  title       = {Differential Inclusions and Viability Theory},
  author      = {Aubin, Jean-Pierre and Cellina, Arrigo},
  year        = {1982},
  number      = {WP-82-051},
  institution = {International Institute for Applied Systems Analysis (IIASA)}
}

@article{Brezis1968,
  author    = {H. Brézis},
  title     = {Equations et inéquations non linéaires dans les espaces vectoriels en dualité},
  journal   = {Annales de l'Institut Fourier},
  volume    = {18},
  number    = {1},
  pages     = {115--175},
  year      = {1968}
}

@book{Brezis1973,
  author    = {H. Brézis},
  title     = {Opérateurs maximaux monotones et semi-groupes de contractions dans les espaces de Hilbert},
  publisher = {North-Holland},
  year      = {1973}
}

@book{Filippov1964,
  author    = {A. F. Filippov},
  title     = {Differential Equations with Discontinuous Right-Hand Sides},
  publisher = {American Mathematical Society},
  year      = {1964}
}

@article{Utkin1977, author = {V. I. Utkin}, title = {Variable Structure Systems with Sliding Modes}, journal = {IEEE Transactions on Automatic Control}, volume = {22}, number = {2}, pages = {212--222}, year = {1977} }

@article{Roxin1966,
  author    = {E. Roxin},
  title     = {On Finite Stability in Control Systems},
  journal   = {SIAM Journal on Control},
  volume    = {4},
  number    = {3},
  pages     = {425--431},
  year      = {1966}
}

@article{BhatBernstein2000,
  author    = {S. P. Bhat and D. S. Bernstein},
  title     = {Finite-Time Stability of Continuous Autonomous Systems},
  journal   = {SIAM Journal on Control and Optimization},
  volume    = {38},
  number    = {3},
  pages     = {751--766},
  year      = {2000}
}

@incollection{diaz2005special,
  title={Special finite time extinction in nonlinear evolution systems: dynamic boundary conditions and Coulomb friction type problems},
  author={D{\'\i}az, JI},
  booktitle={Nonlinear Elliptic and Parabolic Problems: A Special Tribute to the Work of Herbert Amann},
  pages={71--97},
  year={2005},
  publisher={Springer}
}

@article{levaggi2007regularization,
  title={On the regularization of sliding modes},
  author={Levaggi, Laura and Villa, Silvia},
  journal={SIAM Journal on Optimization},
  volume={18},
  number={3},
  pages={878--894},
  year={2007},
  publisher={SIAM}
}

@article{levaggi2013existence,
  title={Existence of sliding motions for nonlinear evolution equations in Banach spaces},
  author={Levaggi, Laura},
  journal={Discrete Contin. Dynam. Systems, Supplement},
  pages={477--487},
  year={2013}
}

@article{baji2007asymptotics,
  title={Asymptotics for some nonlinear damped wave equation: finite time convergence versus exponential decay results},
  author={Baji, B and Cabot, A and D{\'\i}az, JI},
 journal={Annales de l'Institut Henri Poincar{\'e} C, Analyse non lin{\'e}aire},
  volume={24},
  pages={1009--1028},
  year={2007},
  organization={Elsevier}
}

@article{dieci2013regularizing,
  title={Regularizing Piecewise Smooth Differential Systems: Co-Dimension Discontinuity Surface},
  author={Dieci, Luca and Guglielmi, Nicola},
  journal={Journal of Dynamics and Differential Equations},
  volume={25},
  number={1},
  pages={71--94},
  year={2013},
  publisher={Springer}
}

@article{jeffrey2014dynamics,
  title={Dynamics at a switching intersection: Hierarchy, isonomy, and multiple sliding},
  author={Jeffrey, Mike R},
  journal={SIAM Journal on Applied Dynamical Systems},
  volume={13},
  number={3},
  pages={1082--1105},
  year={2014},
  publisher={SIAM}
}

@article{danca2010uniqueness,
  title={On the uniqueness of solutions to a class of discontinuous dynamical systems},
  author={Danca, Marius-F},
  journal={Nonlinear analysis: real world applications},
  volume={11},
  number={3},
  pages={1402--1412},
  year={2010},
  publisher={Elsevier}
}

@article{ball1977strongly,
  title={Strongly continuous semigroups, weak solutions, and the variation of constants formula},
  author={Ball, John M},
  journal={Proceedings of the American Mathematical Society},
  volume={63},
  number={2},
  pages={370--373},
  year={1977}
}

@article{levaggi2004high,
  title={High-gain feedback and sliding modes in infinite dimensional systems},
  author={Levaggi, Laura},
  journal={Control and Cybernetics},
  volume={33},
  number={1},
  pages={33--50},
  year={2004}
}

@article{levaggi2002infinite, title={Infinite dimensional systems’ sliding motions}, author={Levaggi, Laura}, journal={European Journal of Control}, volume={8}, number={6}, pages={508--516}, year={2002}, publisher={Elsevier} }

@article{balogoun2025sliding, title={Sliding mode control for a class of linear infinite-dimensional systems}, author={Balogoun, Isma{\"\i}la and Marx, Swann and Plestan, Franck}, journal={IEEE Transactions on Automatic Control}, year={2025}, publisher={IEEE} }

@book{orlov2020nonsmooth, title={Nonsmooth {Lyapunov} analysis in finite and infinite dimensions}, author={Orlov, Yury}, year={2020}, publisher={Springer} }

@article{orlov2002discontinuous, title={Discontinuous unit feedback control of uncertain infinite-dimensional systems}, author={Orlov, Yuri V}, journal={IEEE transactions on automatic control}, volume={45}, number={5}, pages={834--843}, year={2002}, publisher={IEEE} }

@article{utkin1990control, title={Control of infinite-dimensional plants}, author={Utkin, VI}, journal={Deterministic Control of Uncertain Systems}, number={40}, pages={351}, year={1990}, publisher={Institution of Engineering \& Technology} }

@article{orlov1987sliding, title={Sliding mode control in indefinite-dimensional systems}, author={Orlov, Yu V and Utkin, Vadim I}, journal={Automatica}, volume={23}, number={6}, pages={753--757}, year={1987}, publisher={Elsevier} }

@article{balogoun2022super, title={Super-twisting sliding mode control for the stabilization of a linear hyperbolic system}, author={Balogoun, Isma{\"\i}la and Marx, Swann and Liard, Thibault and Plestan, Franck}, journal={IEEE Control Systems Letters}, volume={7}, pages={1--6}, year={2022}, publisher={IEEE} }

@article{orlov2004robust, title={Robust stabilization of infinite-dimensional systems using sliding-mode output feedback control}, author={Orlov, Yury and Lou, Yiming and Christofides*, Panagiotis D}, journal={International Journal of Control}, volume={77}, number={12}, pages={1115--1136}, year={2004}, publisher={Taylor \& Francis} }

@article{colli2019sliding, title={Sliding mode control for a phase field system related to tumor growth}, author={Colli, Pierluigi and Gilardi, Gianni and Marinoschi, Gabriela and Rocca, Elisabetta}, journal={Applied Mathematics \& Optimization}, volume={79}, number={3}, pages={647--670}, year={2019}, publisher={Springer} }

@article{hajek1979discontinuous,
  title={Discontinuous differential equations, {I}},
  author={H{\'a}jek, Otomar},
  journal={Journal of Differential Equations},
  volume={32},
  number={2},
  pages={149--170},
  year={1979},
  publisher={Elsevier}
}

@inproceedings{brezis1974monotone,
  title={Monotone operators, nonlinear semigroups and applications},
  author={Brezis, Haim},
  booktitle={Proceedings of the International Congress of Mathematicians (Vancouver, BC, 1974)},
  volume={2},
  pages={249--255},
  year={1974}
}

@article{komura1967nonlinear,
  title={Nonlinear semi-groups in {Hilbert} space},
  author={Komura, Yukio},
  journal={Journal of the Mathematical Society of Japan},
  volume={19},
  number={4},
  pages={493--507},
  year={1967},
  publisher={The Mathematical Society of Japan}
}

@book{cross1998multivalued,
  title={Multivalued linear operators},
  author={Cross, Ronald},
  volume={213},
  year={1998},
  publisher={CRC Press}
}

@book{deimling2013nonlinear,
  title={Nonlinear functional analysis},
  author={Deimling, Klaus},
  year={2013},
  publisher={Springer Science \& Business Media}
}

@article{crandall1969semi,
  title={Semi-groups of nonlinear contractions and dissipative sets},
  author={Crandall, Michael G and Pazy, Amnon},
  journal={Journal of functional analysis},
  volume={3},
  number={3},
  pages={376--418},
  year={1969},
  publisher={Academic Press}
}

@article{amaliki_2024,
  title={Stabilization in finite time of a class of evolution equations under multiplicative or additive controls},
  author={Amaliki, Younes and Ouzahra, Mohamed},
  journal={Journal of Control and Decision},
  pages={1--11},
  year={2025},
  publisher={Taylor \& Francis}
}

@article{polyakov2018homogeneous,
  title={On homogeneous finite-time control for linear evolution equation in {Hilbert} space},
  author={Polyakov, Andrey and Coron, Jean-Michel and Rosier, Lionel},
  journal={IEEE Transactions on Automatic Control},
  volume={63},
  number={9},
  pages={3143--3150},
  year={2018},
  publisher={IEEE}
}

@book{orlov2009discontinuous,
  title={Discontinuous systems: {L}yapunov analysis and robust synthesis under uncertainty conditions},
  author={Orlov, Yury V},
  year={2009},
  publisher={Springer}
}

@book{vrabie1995compactness,
  title={Compactness methods for nonlinear evolutions},
  author={Vrabie, Ioan I},
  volume={75},
  year={1995},
  publisher={CRC Press}
}

@article{carjua2001viability,
  title={Viability for semilinear differential inclusions via the weak sequential tangency condition},
  author={C{\^a}rj{\u{a}}, Ovidiu and Vrabie, Ioan I},
  journal={Journal of mathematical analysis and applications},
  volume={262},
  number={1},
  pages={24--38},
  year={2001},
  publisher={Elsevier}
}

@article{carjua1997some,
  title={Some new viability results for semilinear differential inclusions},
  author={C{\^a}rj{\u{a}}, Ovidiu and Vrabie, Ioan I},
  journal={Nonlinear Differential Equations and Applications NoDEA},
  volume={4},
  number={3},
  pages={401--424},
  year={1997},
  publisher={Springer}
}

@misc{balogoun2022slidingmodecontrolclass,
      title={Sliding mode control for a class of linear infinite-dimensional systems}, 
      author={Ismaïla Balogoun and Swann Marx and Franck Plestan},
      year={2022},
      eprint={2210.13465},
      archivePrefix={arXiv},
      primaryClass={math.AP},
      url={https://arxiv.org/abs/2210.13465}, 
}

@article{liard2022boundary,
  title={Boundary sliding mode control of a system of linear hyperbolic equations: A    {Lyapunov}     approach},
  author={Liard, Thibault and Balogoun, Isma{\"\i}la and Marx, Swann and Plestan, Franck},
  journal={Automatica},
  volume={135},
  pages={109964},
  year={2022},
  publisher={Elsevier}
}

@incollection{brezis1971monotonicity,
  title={Monotonicity methods in {Hilbert} spaces and some applications to nonlinear partial differential equations},
  author={Br{\'e}zis, Ha{\"\i}m},
  booktitle={Contributions to nonlinear functional analysis},
  pages={101--156},
  year={1971},
  publisher={Elsevier}
}

@book{brezis1973ope,
  title={Op\'erateurs maximaux monotones et semi-groupes de contractions dans les espaces de {Hilbert}},
  author={Brezis, Haim},
  year={1973},
  publisher={North Holland}
}

@book{pazy2012semigroups,
  title={Semigroups of linear operators and applications to partial differential equations},
  author={Pazy, Amnon},
  volume={44},
  year={2012},
  publisher={Springer Science \& Business Media}
}

@article{cheng2011sliding,
  title={Sliding mode boundary control of a parabolic {PDE} system with parameter variations and boundary uncertainties},
  author={Cheng, Meng-Bi and Radisavljevic, Verica and Su, Wu-Chung},
  journal={Automatica},
  volume={47},
  number={2},
  pages={381--387},
  year={2011},
  publisher={Elsevier}
}

@article{pisano2011tracking,
  title={Tracking control of the uncertain heat and wave equation via power-fractional and sliding-mode techniques},
  author={Pisano, Alessandro and Orlov, Yury and Usai, Elio},
  journal={SIAM Journal on Control and Optimization},
  volume={49},
  number={2},
  pages={363--382},
  year={2011},
  publisher={SIAM}
}

@article{guzman2025rapid,
  title={Rapid stabilization for a wave equation with boundary disturbance},
  author={Guzm{\'a}n, Patricio and Huerta, Agust{\'\i}n and Parada, Hugo},
  journal={arXiv preprint arXiv:2510.04893},
  year={2025}
}

@article{fxt:Polyakov:2012,
  author={Polyakov, Andrey},
  journal={IEEE Transactions on Automatic Control}, 
  title={Nonlinear Feedback Design for Fixed-Time Stabilization of Linear Control Systems}, 
  year={2012},
  volume={57},
  number={8},
  pages={2106--2110},}

@article{polyakov2021input,
  title={Input-to-State Stability of homogeneous infinite dimensional systems with locally {L}ipschitz nonlinearities},
  author={Polyakov, Andrey},
  journal={Automatica},
  volume={129},
  pages={109615},
  year={2021},
  publisher={Elsevier}
}

\end{document}